\documentclass[twocolumn]{aastex63}

\usepackage{hyperref}
\usepackage{verbatim}
\received{2022 June}
\usepackage{amsmath}
\usepackage{amsfonts}
\shorttitle{Black Widow Population}
\shortauthors{Kandel \& Romani}
\defcitealias{linares2018peering}{LSC18}
\defcitealias{romani2015keck}{RGFK15}
\defcitealias{van2011evidence}{VBK11}

\begin{document}

\title{An Optical Study of the Black Widow Population}

\correspondingauthor{R.W. Romanil}
\email{rwr@stanford.edu}

\author[0000-0002-5402-3107]{D. Kandel}
\affil{Department  of  Physics,  Stanford  University,  Stanford,  CA, 94305, USA}

\author[0000-0001-6711-3286]{Roger W. Romani}
\affil{Department  of  Physics,  Stanford  University,  Stanford,  CA, 94305, USA}
\begin{abstract}
The optical study of the heated substellar companions of `Black Widow' (BW) millisecond pulsars (MSP) provides unique information on the MSP particle and radiation output and on the neutron star mass. Here we present analysis of optical photometry and spectroscopy of a set of relatively bright BWs, many newly discovered in association with {\it Fermi} $\gamma$-ray sources. Interpreting the optical data requires sophisticated models of the companion heating. We provide a uniform analysis, selecting the preferred heating model and reporting on the companion masses and radii, the pulsar heating power and neutron star mass. The substellar companions are substantially degenerate, with average densities $15-30\times$ Solar, but are inflated above their zero temperature radii. We find evidence that the most extreme recycled BW pulsars have both large $>0.8M_\odot$ accreted mass and low $<10^8$G magnetic fields. Examining a set of heavy BWs, we infer that neutron star masses larger than $2.19 M_\odot$ ($1\sigma$ confidence) or $2.08 M_\odot$ ($3\sigma$ confidence) are required; these bounds exclude all but the stiffest equations of state in standard tabulations.
\end{abstract}


\section{Introduction} \label{sec:intro}
Black widows (BWs) with sub-stellar companions and redbacks (RBs) with low-mass star companions, together known as  `spiders',  are binary systems of MSP in tight $<1$\,d orbits, with the companion heated and evaporated by the pulsar spindown power. Since the discovery of BW PSR J1959+2048's companion\citep{1988Natur.333..237F,1988Natur.334..504K}, it has been known that the heating pattern and the resulting light curve are important probes of the binary geometry (\citealt{djorgovski1988photometry}, \citealt{callanan1995orbital}) and the mass of the MSP (\citealt{aldcroft1992spectroscopy}, \citealt{van2011evidence}). These pulsars are difficult to discover in the radio band, due to scattering and obscuration by the evaporation wind. The pulsars' peak photon output is in the penetrating GeV gamma-rays and with the advent of the {\it Fermi} LAT sky survey and attendant follow-up searches, the number of known `spiders' has greatly increased. We now have detailed optical light curves of many of these objects \citep{draghis2019multiband}. Photometric data and its modeling provide an important probe of the physics of these systems and can constrain the size and heating of the companion. Most importantly, such modeling constrains the binary inclination $i$, which together with spectroscopic measurements can help determine the pulsar and companion masses.

Traditional modeling of these spider systems assumes direct pulsar irradiation of the companion, which then spontaneously re-emits thermal radiation. For such models, the optical light curve is symmetric with its maximum at pulsar inferior conjunction. While some objects show optical modulation broadly consistent with this picture, many light curves, especially those with higher S/N, show significant peak asymmetries \citep{stappers2001intrinsic} and phase-shifts \citep{schroeder2014observations}. Past work has discussed the possible physical origin of such asymmetries, including companion irradiation by an asymmetric intrabinary shock (IBS; \citealt{2016ApJ...828....7R}), IBS particles ducting along magnetic field lines to companion poles \citep{2017ApJ...845...42S}, heat transfer from a global wind (\citealt{kandel2020atmospheric}, hereafter KR20) or general surface diffusion \citep{voisin2020model}. As described in \citet{kandel2020atmospheric}, and \citet{romani2021psr},  properly accounting for these effects is important to get an unbiased estimate of the NS mass. We caution that, in many cases, masses based on simple direct heating modeling will not be accurate.

In this paper, we discuss ten BW systems, presenting uniform LC modeling of nine for which we have photometric data. Six out of these also have radial velocity data, which gives us an opportunity to estimate the system masses. We show that BW systems host heavy neutron stars, and by combining the inferred neutron star masses, we can put a strong lower bound on the maximal neutron star mass, appreciably above the values inferred from radio Shapiro delay measurements. This should have significant ramifications for the equation of state of dense matter.

\begin{deluxetable*}{lcccccccc}
\tabletypesize{\footnotesize}
\tablewidth{0pt}
\tablecaption{Summary of BW parameters and photometric observations\label{tab:objects_summary}}
\tablehead{\colhead{Name} & \colhead{$P_s$ (ms)} & \colhead{$P_b$ (hr)} & \colhead{$\dot{E}(10^{34}$erg/s)} & \colhead{$x_1$(lt-s)} & \colhead{$A_V$(mag)} & \colhead{$B\,(10^8$G)}& \colhead{Bands}&  \colhead{Ref.}
}
\startdata
J$0023+0923$& 3.05 &3.33 & 1.60 & 0.035 &0.382 & 1.90&$gri$ & \cite{draghis2019multiband}\cr
J$0636+5128$&2.87 &1.60 & 0.58 & 0.009 & 0.218 & 1.01& $grizHK$ & \cite{draghis2018psr} \cr
J$0952-0607$ & 1.41 & 6.42 & 6.65 &0.063& 0.137 & 0.82& $ugriz$ & \cite{romani2022psr}\cr
J$1301+0833$&1.84 & 6.48 & 6.65 & 0.078 & 0.082 & 1.41&$griz$ & \cite{draghis2019multiband}\cr
J$1311-3430$&2.56 &1.56 & 5.00 & 0.011 & 0.137 &  2.36 & $ugriz$ & \cite{romani2015spectroscopic}\cr
J$1653-0158$& 1.97 & 1.25&1.20 &0.011 & 0.710 & 0.68&  $u'g'r'i'$ & \cite{nieder2020discovery}\cr
J$1810+1744$& 1.66 & 3.56 & 3.97& 0.095 & 0.390 & 0.88 &  $ugriz$& \cite{romani2021psr}\cr
J$1959+2048$& 1.61 & 9.17 & 16.0 & 0.089 & 0.600 & 1.66 &  $BVRIK$  & \cite{reynolds2007light}\cr
J$2051-0827$& 4.51 & 2.37 & 0.55 & 0.045 & 0.296 & 2.42 &   $ugriz$  & \cite{2022arXiv220809249D}\cr
J$2052+1219$&1.99 & 2.75 & 3.37& 0.061& 0.328 & 1.17 &  $gri$ & \cite{draghis2019multiband}\cr
\enddata
 \label{table:objects_summary}
\end{deluxetable*}

\section{Observations}
Table \ref{tab:objects_summary} summarises some basic properties of the ten BWs in our population study. It also shows the photometric bands used; the acquisition and processing of these data have been described in earlier papers. Photometry of five systems -- J0023+0923, J0636+5128, J1301+0833, J1959+2048 and J2052+1219 -- is described in \cite{draghis2019multiband} and references therein. Optical photometry and spectroscopy of J0952$-$0607 is described in \citet{2022ApJ...934L..17R}. For J1311$-$3430, we use synthesized photometry from LRIS spectra, as described in \cite{romani2015spectroscopic}.  For J1653$-$0158, we use ULTRACAM data from \cite{nieder2020discovery}. Optical observations of J1810+1744 are described in \cite{romani2021psr}. Photometry and fitting of PSR J2051$-$0827 are presented in \citet{2022arXiv220809249D}.

\section{Photometric Fitting}

\begin{deluxetable*}{lccccccccccc}
\tabletypesize{\footnotesize}
\tablewidth{0pt}
\tablecaption{Light Curve Fit Results\label{tab:final_fit}}
\tablehead{
\colhead{Name} & \colhead{$i\, (\mathrm{ deg})$} & \colhead{$f$} & \colhead{$T_N\,(\mathrm{K})$} & \colhead{$L_{\mathrm{H}}\,(10^{33}\,\mathrm{erg/s})$} & \colhead{$d\,({\rm kpc})$} & \colhead{$\theta_{\rm hs}\, (\mathrm{deg})$} & \colhead{$\phi_{\rm hs}\, (\mathrm{deg})$} & \colhead{$\mathcal{A}_{\rm hs}$} &\colhead{$\sigma_{\rm hs}\, (\mathrm{deg})$} & \colhead{$\epsilon$} & \colhead{$\chi^2/{\rm DoF}$}}
\rotate
\startdata
J$0023+0923$& $74.1^{+8.8}_{-12}$&$0.87^{+0.04}_{-0.05}$ &$2992^{+94}_{-92}$ &$1.22^{+0.30}_{-0.30}$ &$0.59^{+1.39}_{-0.39}$ &$353.0^{+2.7}_{-5.5}$ &$-22.6^{+75.8}_{-42.1}$ &$0.62^{+0.16}_{-0.15}$  & $11.4^{+3.8}_{-2.6}$&$-$&$105/53$\\
\\
J$0636+5128$&$34.2^{+2.9}_{-2.5}$ &$0.88\pm 0.02$ & $2284^{+100}_{-100}$&$1.9^{+0.3}_{-0.3}$ & $0.63^{+1.44}_{-0.43}$ &$258.2^{+11.3}_{-12.4}$  &$66.1^{+6.9}_{-8.9}$ & $1.0^{+0.10}_{-0.13}$&$11.0^{+3.0}_{-2.2}$ &- &$235/155$\\
\\
J$0952-0607$&$59.8^{+2.0}_{-1.9}$ & $0.79\pm 0.01$ & $3085^{+85}_{-80}$&$3.81^{+0.46}_{-0.43}$& $6.26^{+0.36}_{-0.40}$&-&-&-&-&-&286/287\\
\\
J$1301+0833$&$46.6^{+2.5}_{-2.2}$ &$0.61\pm 0.06$ &$2288^{+80}_{-84}$ &$4.7^{+0.5}_{-0.5}$ &$1.77^{+0.10}_{-0.11}$ &- &- &- &- &- &157/121 \\
\\
J$1311-3430$& $68.7^{+2.1}_{-2.0}$&$0.99$ &$821
^{+959}_{-402}$ &$228^{+56}_{-44}$ &$3.01\pm 0.15$ &$119.8^{+4.3}_{-4.3}$ &$27.5^{+15.7}_{-17.5}$ &$11.2^{+10.6}_{-5.9}$  & $16.8^{+3.4}_{-3.1}$&$-0.072^{+0.001}_{-0.001}$&$74/84$\\
\\
J$1653-0158$&$72.8^{+4.0}_{-4.0}$ &$0.84^{+0.02}_{-0.02}$ &$2253^{+499}_{-389}$ &$4.3^{+0.4}_{-0.6}$ &$0.87^{+0.07}_{-0.08}$ & $287.1^{+5.7}_{-5.2}$& $-48.1^{+6.9}_{-5.2}$& $1.23^{+0.47}_{-0.36}$& $21.8^{+3.4}_{-3.2}$& -& 1296/941\\
\\
J$1810+1744$&$66.3^{+0.5}_{-0.5}$ &0.99 &$3101^{+77}_{-94}$ &$41.8^{+3.7}_{-4.8}$ &$2.86\pm 0.13$ & $86.9^{+1.1}_{-0.8}$&$23.0^{+3.2}_{-2.6}$ &$0.86\pm 0.23$ &$23.4^{+1.2}_{-1.1}$  &$-0.19\pm 0.02$ &229/168 \\
\\
J$1959+2048^*$ &$85.3^{+2.9}_{-1.2}$& $0.74^{+0.03}_{-0.04}$ & $2710^{+25}_{-23}$ & $42.7^{+2.7}_{-2.2}$ & $2.25\pm 0.11$ & -&-&-&-&-&128/88\\
\\
J$2051-0837^*$&$55.9^{+4.8}_{-4.1}$ &$0.88^{+-.02}_{-0.02}$ &$2750^{+120}_{-150}$ &$1.7^{+0.3}_{-0.3}$ &$2.48^{+0.39}_{-0.38}$ &- &- &- &- &- &1288/1341 \\
\\
J$2052+1219$&$59.3^{+2.3}_{-1.8}$ &$0.99$ &$2710^{+332}_{-1218}$ &$11.8^{+3.5}_{-3.4}$ &$5.57^{+0.47}_{-0.55}$ &- &- &- &- &$0.03\pm 0.01$ &153/80 \\
\\
\enddata
 \label{table:final_fit}
\end{deluxetable*}

Our fits are performed with an outgrowth of the {\tt ICARUS} light-curve model \citep{breton2012koi} with additions described by \cite{2020ApJ...903...39K}. An additional update replaces the simplified limb-darkening laws in the base code with the more detailed limb-darkening coefficients computed by \cite{claret2011gravity} for two models, the ATLAS and PHOENIX atmospheres. We generally find that the ATLAS coefficients perform better. Note that gravity- and limb-darkening serve to rescale the local fluxes; to monitor the heating budget we take care to integrate the emergent flux to determine the total companion heating $L_H$, subtracting the thermal base emission (characterized by the night-side temperature $T_N$). In all fits, we explore a simple direct heating (DH) model, a wind model (WH), and a hotspot model (HS) and perform model selection using  Akaike Information Criterion \citep[AIC,][]{1974ITAC...19..716A}. In general, asymmetry of the light-curve maximum and bluer colors to one side of the peak indicate can the presence of a hotspot, whereas peak broadening and a flux gradient across the maximum tend to indicate heat advection along latitude lines from global winds.

In our fits, we choose the gravity darkening coefficient to be $\beta_{\rm D}=$0.08 for companions with low front-side temperature ($T\lesssim 10,000$\,K) and 0.25 for front-side temperature $T\gtrsim 10,000$\,K. To check whether radiative conditions apply, for imtermediate front-side temperature $5,000\lesssim T\lesssim 9,000$ K, we tried fitting the gravity darkening coefficient and found that $\beta_{\rm D}$ of 0.08 was strongly preferred over 0.25; for such cases, $\beta_{\rm D}=0.08$ was fixed during model fitting.

The observed lightcurves are also affected by the intervening extinction $A_V$. For many of our objects, we adopt the extinction estimates from the three-dimensional dust model of \cite{green20193d} available as a query at \url{argonaut.skymaps.info}, using the {\tt Bayestar2019} model and $A_V=2.73\,E(g-r)$ values at the estimated distance. For southern sources, we take the total extinction column through the Galaxy from the NASA Extragalactic Database \citep{schlafly2011measuring} as an upper limit.

For the light curve modeling, the principal geometrical and physical fit parameters are the inclination of the binary system $i$, the Roche lobe filling factor $f$ (defined as the ratio of the companion nose radius to the radius of $L_1$ point), the base temperature of the star $T_N$, the pulsar irradiation power $L_H$, and the distance to the binary system $d$ in kpc. For HS models, we also fit the Gaussian spot amplitude $\mathcal{A}_{\rm hs}$ (the multiplicative increase to the local temperature), radial size $r_{\rm hs}$, and angular position $\theta_{\rm hs}$ (measured from the sub-pulsar point on the companion), $\phi_{\rm hs}$ [measured from the orbital plane, 0 toward the dusk (East) terminator of the companion and $\pi/2$ toward the North (spin) pole]. For WH models, we fit for the wind parameter $\epsilon=\tau_{rad}/\tau_{\rm adv}$, the ratio of the radiation cooling and advection timescales. For $i$, a uniform prior in $\cos\,i$ was applied; for $L_H$, a uniform logarithmic prior was applied, whereas for all other parameters a uniform prior was used. For all our fits, we explore models with increasing model complexity: a direct heating (DH) model, DH+wind heating (WH), DH+ hot spot(HS), and DH+WH+HS. To penalize increased model complexity, the final model selection is based on Akaike Information Criterion.

Photometric fitting gives us constraints on heating parameters as well as binary geometry. This can be combined with center-of-mass velocity obtained from radial velocity (RV) fitting to estimate the NS and companion masses. In general, the observed RV amplitude is different from the center-of-mass (CoM) radial velocity amplitude $K_{\rm CoM}$ because strong pulsar irradiation shifts the photometric center of the secondary star (``center of light" CoL) away from its CoM. Moreover, different absorption lines have strengths varying with the surface temperature; thus the CoL shift is different for each spectral line. The temperature sensitivity of the line profile can be characterized by monitoring its equivalent-width $EW(T)$ variation across the companion surface. Therefore, we model the CoL RVs for a given $K_{\rm CoM}$, by averaging up the radial velocities over the companion surface elements, weighted by flux and temperature-dependent equivalent-width $EW(T)$ of the lines that dominate the radial velocity template. We fit these predicted CoL RVs to the cross-correlation velocities, estimating $K_{\rm CoM}$ and hence the binary masses. Although we do not perform a simultaneous photometry fit, we do marginalize the spectroscopic fit over the geometrical parameters from the end of the photometric Markov Chain Monte Carlo chains, sampling $\sim 2\sigma$ uncertainties. Thus, the mass errors do include all uncertainties in the model fitting, spectroscopic and photometric.

Below, we describe photometric and RV fitting of individual BW objects. Our population study includes J0952-0607 and J1810+1744, but since these objects were already fitted with the most updated prescription in \cite{romani2022psr} and \cite{romani2021psr}, we skip the modeling details and present only the fit results. The values for J2051$-$0827 are from \citet{2022arXiv220809249D}.

\subsection*{PSR J0023+0923}
Lacking radial velocity measurements, we do not have enough constraints for a detailed NS mass estimate, so we fix at  $M_{\rm NS}=1.5\,M_\odot$. We find that a HS model fits the data the best, with $\chi^2/$DoF $=107/57$ and a poorly constrained $i(^\circ)=74.5^{+9.7}_{-15.5}$. With this inclination, the companion mass is rather small $M_c \sim 0.018 M_\odot\,(M_{\rm NS}/1.5\,M_\odot)^{2/3}$, typical of other BWs. The maximum is relatively well sampled, thus the fill-factor is modestly constrained at $0.87\pm 0.05$, implying a volume-averaged companion radius of $0.10\,R_\odot$. As noted in \cite{draghis2018psr}, BW companions appear to be inflated by the heating and hence we expect companion radius to be larger than the radius $R_{\rm cold}$ of a cold (degenerate) remnant of a stellar core. This heating-induced inflation is discussed in \S4.2. Due to limited photometry near optical minimum, our inclination and hot-spot position estimates have substantial uncertainty. While the hotspot's nose angle $\theta_{\rm hs}$ is near the nose, its phase angle is highly unconstrained. Additional photometry covering the minimum, especially in the $i$ and $z$ bands, would help constrain $T_N$ and $i$, and $u$ band data would help constrain hot-spot parameters. 

\begin{figure*}
\centering
\includegraphics[scale=0.25]{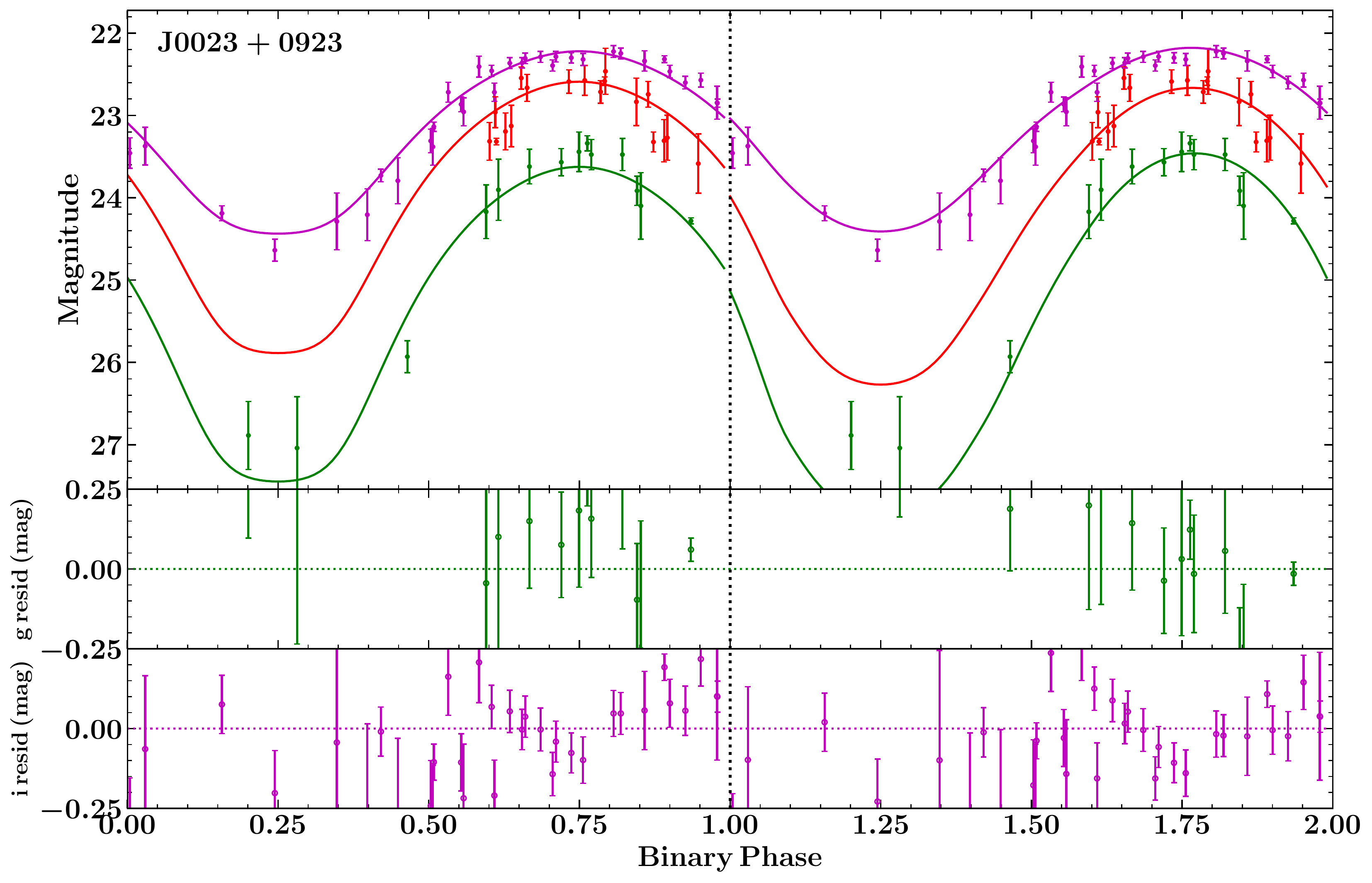}
\includegraphics[scale=0.25]{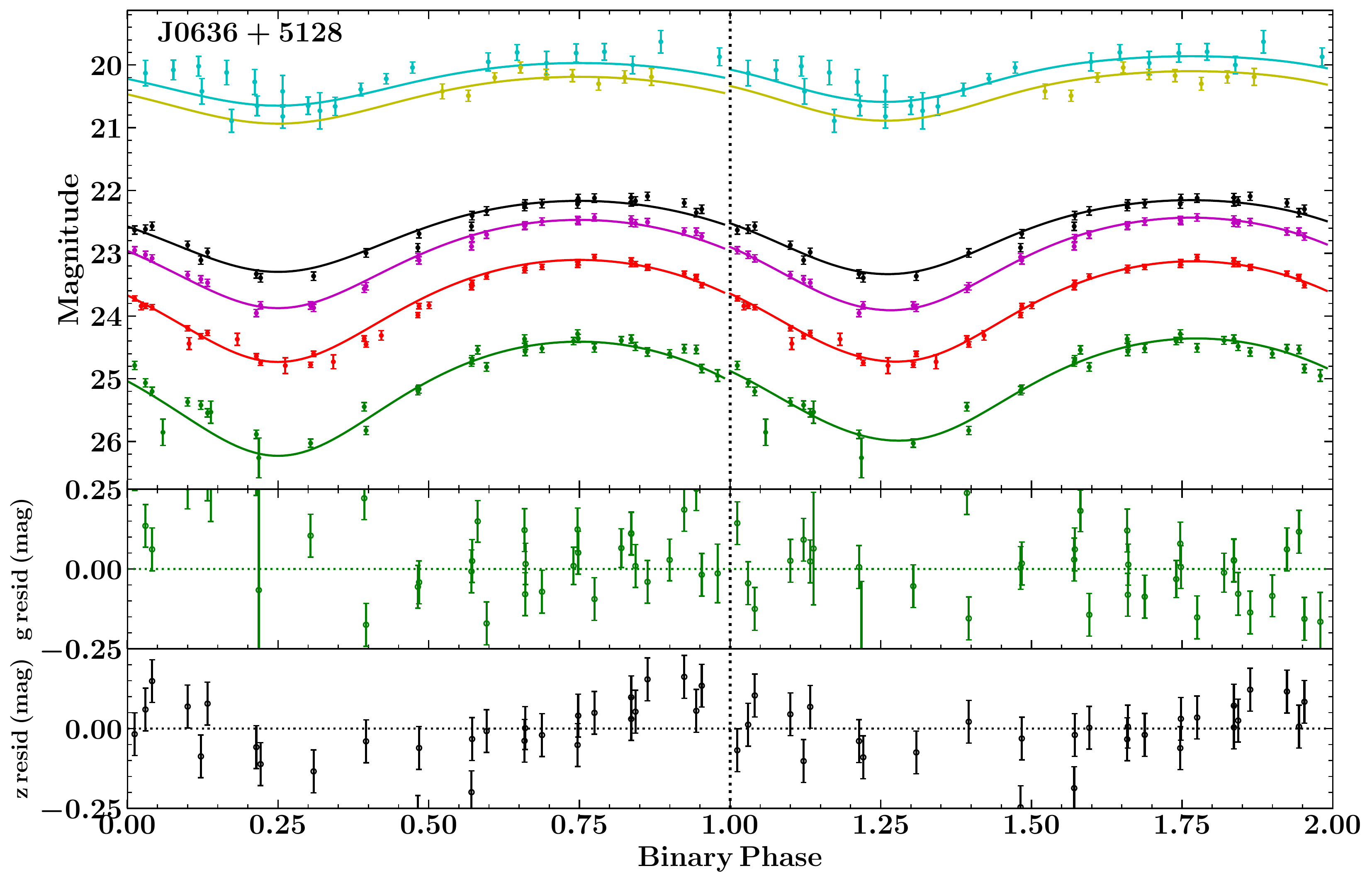}
\includegraphics[scale=0.25]{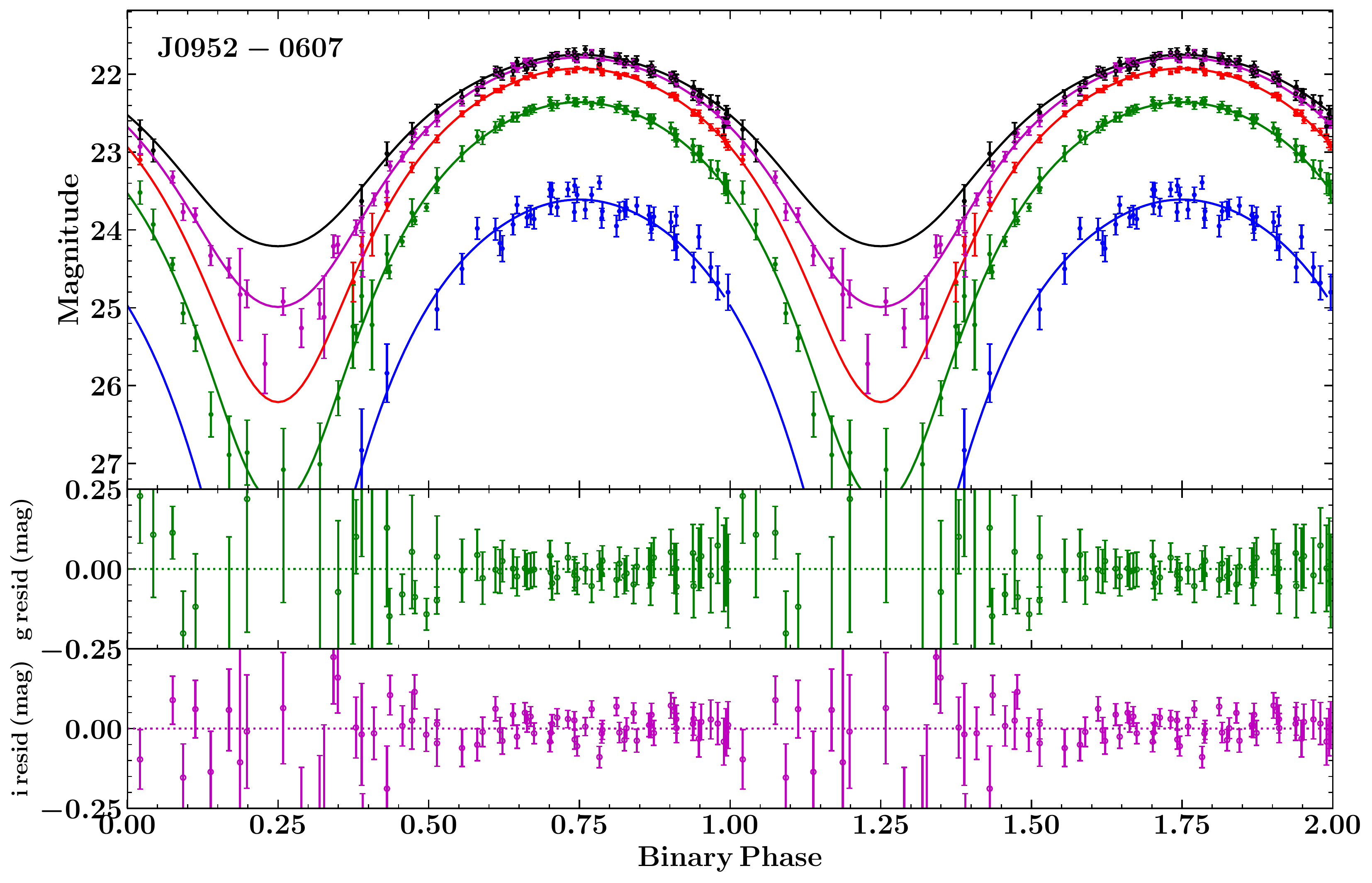}
\includegraphics[scale=0.25]{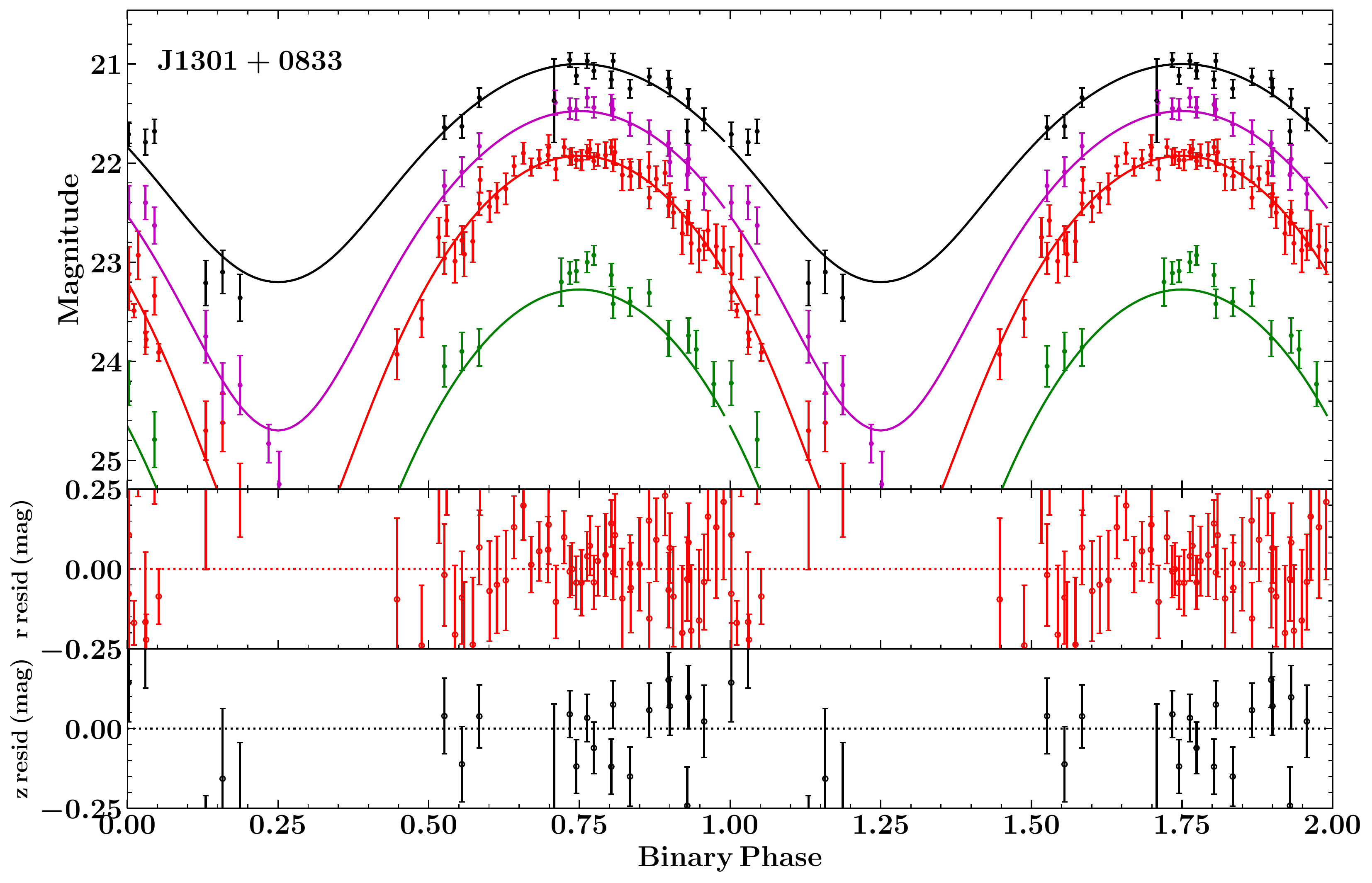}
\caption{Four BW LCs. $\phi_{\rm B}=0$ is defined as pulsar TASC (time of the pulsar ascending node). Two cycles are shown. For J0952 and J1301 direct heating provides the best-fit model and is shown in both cycles. For the others the first cycle shows the best-fit direct heating model, and the second is the best-fit hotspot or wind model. Lower panels show residuals to the model (for a red band and a blue band) of the corresponding cycle -- the model differences are often subtle, best seen in the residuals, but are statistically significant (see text). Colors denote the various photometric bands (blue=$u$, green=$g$ red=$r$, magenta=$i$, black=$z$, yellow=$H$, cyan=$K$.}
\label{fig:4lca}
\end{figure*}

\begin{figure*}
\centering
\includegraphics[scale=0.25]{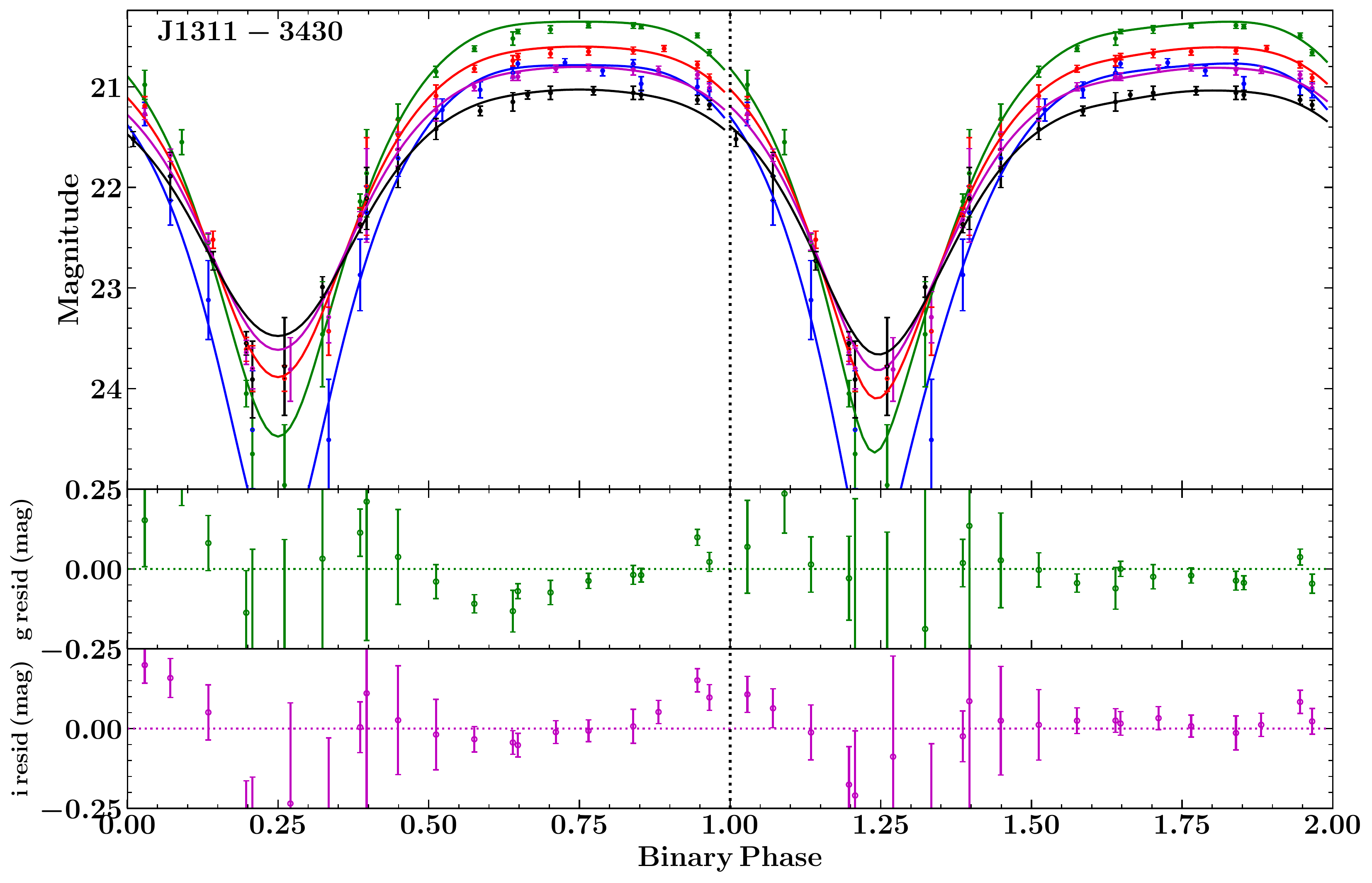}
\includegraphics[scale=0.25]{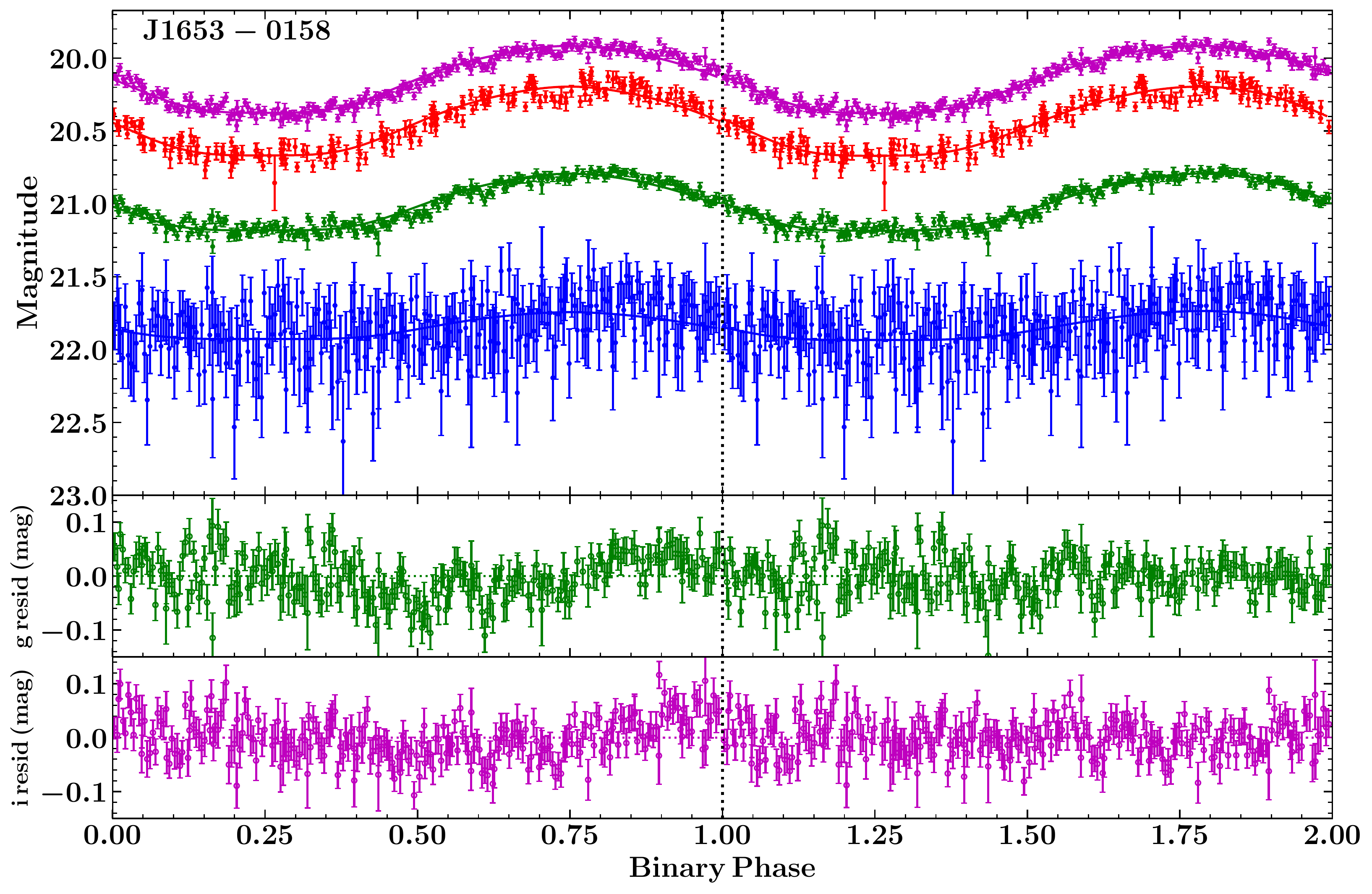}
\includegraphics[scale=0.25]{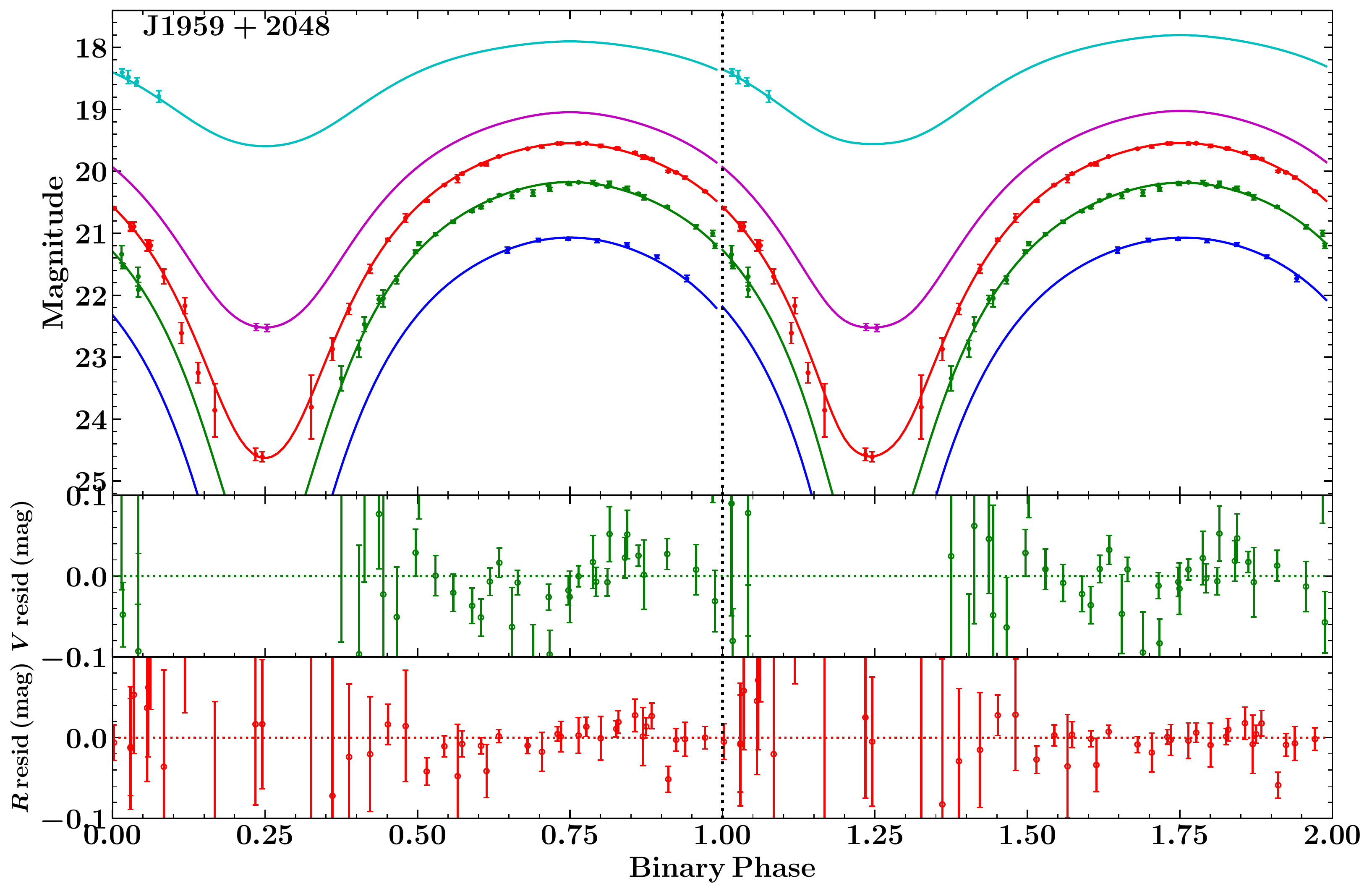}
\includegraphics[scale=0.25]{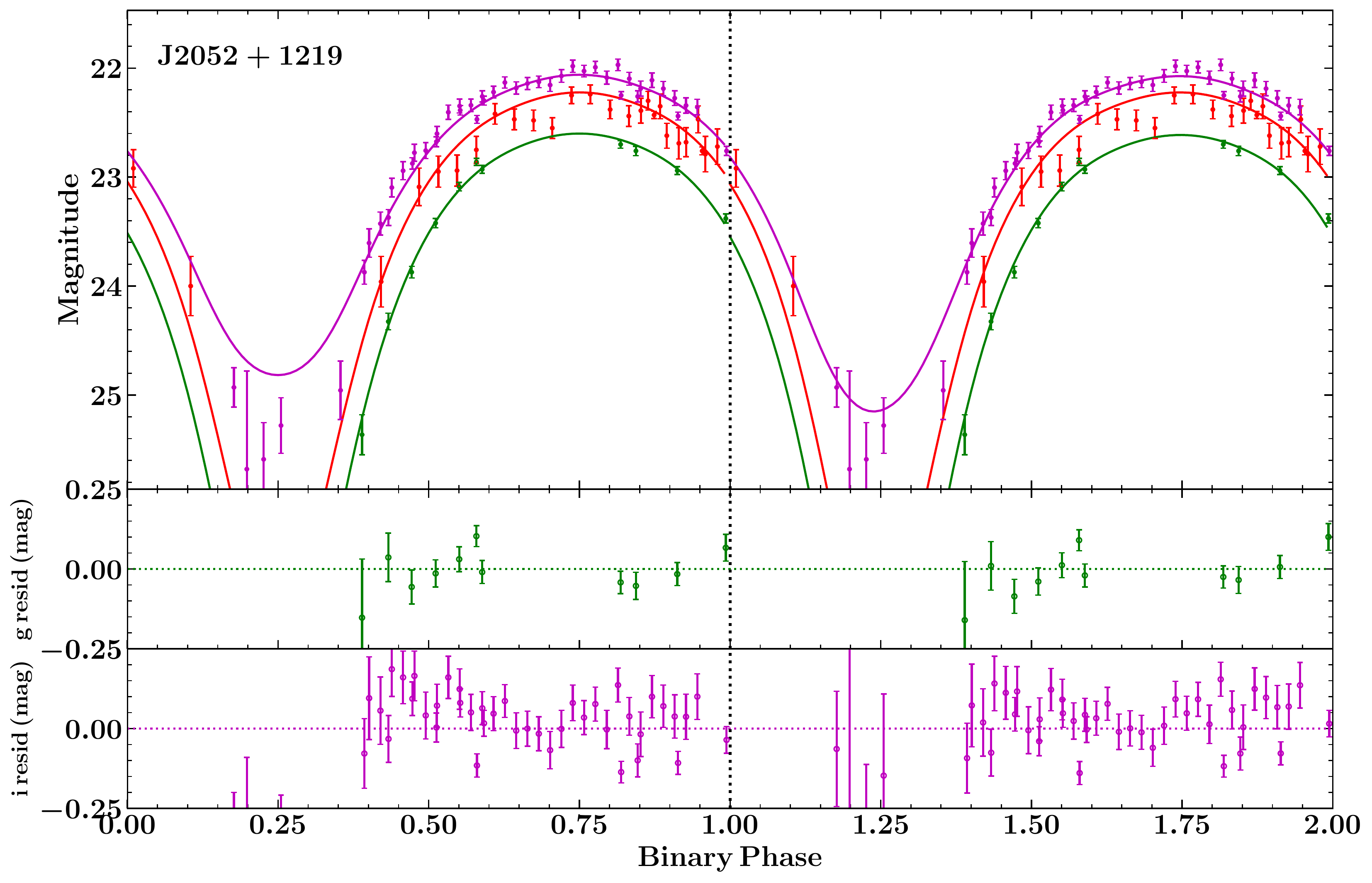}
\caption{As for figure \ref{fig:4lca}, except for J1959+2048, where blue=$B$, green=$V$, red=$R$, magenta=$I$.}
\label{fig:4lcb}
\end{figure*}

\subsection*{PSR J0636+5128}
Fig \ref{fig:4lca} shows the $grizHK$ LC of this object. The LC is relatively shallow for all colors, indicating a low binary inclination. Since radial velocity measurements are unavailable, we cannot fit the NS mass, so we fix at $M_{\rm NS}=1.5\,M_\odot$. The best model is achieved with a $11.0^\circ$ Gaussian hotspot located close to the pole of the `northern' hemisphere, with a substantial temperature excess of 64\%. The direct heating power is $1.9\times 10^{33}\,$erg/s representing about $36\%$ of the spin-down power. With the HS model, the inclination is $i\approx 34.0^\circ$, slightly larger than the DH estimate in \cite{draghis2019multiband}. With this inclination, the companion mass is rather small $M_c \sim 0.01 M_\odot(M_{\rm NS}/1.5\,M_\odot)^{2/3}$, so for a modest $M_{\rm NS}>1.5\,M_\odot$, J0636's has a BW-type companion. The timing parallax \citep{stovall2014green} of J0636 gave a very small distance, but this has been superseded by more extended timing, giving $d=1.1^{+0.6}_{-0.3}$ kpc \citep{arzoumanian2018nanograv}, and our photometric distance fit of $0.63^{+1.44}_{-0.43}$ is in good agreement with this value although the constraint is  rather loose. At this distance, the fill factor of $0.88$ gives a companion radius of $0.095\,R_\odot$.

\subsection*{PSR J1301+0833}
\cite{romani2016psr} found that for J1301, the combination of faint magnitude and  $\sim 4500$ K effective temperature companion requires a distance of $\sim 6$\,kpc for direct full-surface heating of a Roche-lobe filling star. This is dramatically larger than the $d = 1.23$\,kpc implied by the pulsar's $DM=13.2\,{\rm pc\,cm}^{-3}$ \citep{ray2012radio} in the {\tt YMW17} model. We have fit with DH, HS and WH models and find that all prefer a companion that substantially under-fills its Roche lobe at $f\sim 0.61$. This is the smallest fill factor found for any of our BWs, but with this value the fit distance is a plausible 1.77\,kpc. While the HS and WH models do slightly decrease $\chi^2$ (by 10 and 4, respectively), with the extra parameters AIC prefers the basic DH model. All models give a similar $i\approx 46.6^\circ$. This small inclination is quite consistent with the relatively small RV amplitude of \cite{romani2016psr}. At our fit distance the Shklovskii-corrected spin-down power is $\dot{E}_c=\dot{E}I_{45}\left[1-(0.31\pm 0.07)d_{\rm kpc}\right]=3.0\times 10^{34}\,$erg/s for the moment of inertia $10^{45}I_{45}$ g/cm$^2$. Thus, our best-fit $L_H=4.7\times 10^{33}$ erg/s is a modest $16\%$ of ${\dot E}$.

\subsection*{PSR J1311-3430}

J1311 is particularly interesting since \cite{romani2012psr} found that the NS might be very massive. \citet{romani2015spectroscopic} used early versions of several HS models to show that the fit $i$ could range between $64^\circ-85^\circ$, with the NS mass as low as $1.8\,M_\odot$ and as large as $2.7\, M_\odot$. With our improved modeling we can reduce these ranges.

Figure \ref{fig:4lcb} shows the LCs of this object from simultaneous photometry synthesized from Keck LRIS spectra. Simultaneous light curves are essential for this object, since J1311 is known to have intense optical flaring, with detectable flares occurring nearly every orbit and large flares appearing every dozen orbits. With simultaneous colors and multiple orbits we can prune these flares leaving the underlying quiescent thermal flux to be described by the heating model. J1311's LCs show three main features: i) the maximum is particularly flat, indicating strong gravity darkening, ii) the maximum is asymmetrical, especially in bluer colors, iii) a slight gradient is present across maximum. All fits find that the companion is very close to Roche-lobe filling, so we set the fill factor at $f=0.99$. The large fill factor (low surface gravity near the $L_1$ point) and the high average front-side temperature $>10,000$ K should require a large $\beta_{\rm D}$. Indeed, if freed in the fits, $\beta_{\rm D}=0.30\pm 0.06$, consistent with the $\beta_{\rm D}=0.25$ expected for a radiative atmosphere.

In our fitting, the best model invokes both HS + WH, with a sub- Alfv\'{e}nic wind with $\epsilon = -0.07\pm 0.01$, and a hotspot just past the day-night terminator with a large $\approx 12\times$ excess over the local nightside temperature. The resulting binary inclination is quite well constrained at $68.7^\circ\pm 2.1^\circ$. We also find that the base temperature $T_N$ is very low, with samples converging below the lowest temperature limit of 1000K of our spectral library. Such extrapolation to low temperature will likely be imprecise as unmodelled effects such as dust settling and clouds are important for low-temperature atmosphere spectra \citep{husser2013new}. Thus, we consider $T_N$ low, but poorly constrained. Note that the inclination is hardly affected by this uncertainty as the model light curves are dominated by hot surface elements resulting from extreme pulsar irradiation.

Using the DM$=37.84$ pc cm$^{-3}$ \citep{ray2013radio}, \cite{antoniadis2021gaia} estimated $2.43\pm 0.48$ kpc. Our distance estimate of $3.01\pm 0.15$ kpc is consistent with this DM estimate. The best-fit $L_H=2.28\times10^{35}$ erg/s is $\approx 4$ larger than the nominal spin-down power. For J1311, a substantial portion of the flux seems to be from IBS particle precipitation (a strong hot spot) and some leakage from this particle-mediate heating may also contribute to the large $L_H$.

\subsection*{PSR J1653-0158}
The lightcurve shows shallow modulation implying either a low inclination or a strong veiling flux. However, with the companion RV amplitude $K_{CoL}=669\pm7.5\,{\rm km\,s^{-1}}$ we find a large mass function $f(M)=1.60\pm0.05M_\odot$ \citep{romani20142fgl}, so small inclination would lead to an unphysically large NS mass. Noting that the minimum is flat and that the modulation decreases for bluer colors, we infer that a strong blue veiling flux dominates at orbital minimum. This is also visible in the phase-resolved spectra \citep{romani20142fgl}.  Although this veiling flux is likely associated with the IBS, we model it here as a simple power-law with form $f_\nu = f_A (\nu/10^{14}{\rm Hz})^{-p}$, with $f_A\sim 101\pm 10\,\mu$Jy and $p=0.50\pm 0.03$. This flux is fairly constant through the orbit, although there are hints of sharp phase structures in the light curve, e.g. in $r$ and $i$ at $\phi_{\rm B}=0.72$. Any model without such a hard-spectrum component is completely  unacceptable. These fits prefer an $A_V\sim 1.0$ mag slightly higher than, but consistent with the maximum in this direction (which is found for all  $d > 300$pc).

We find that the HS model best fits the data. However, with the fine structure near maximum, the model is not yet fully acceptable  ($\chi^2/DoF \sim 1.38$). More detailed models, including modulated veiling flux from the IBS, may be needed to fully model the light curves. Such modeling would be greatly helped by light curves over an even broader spectral range, with IBS effects increasingly dominant in the UV and low-temperature companion emission better constrained in the IR. With many cycles, we could also gauge the reality (and stability) of the apparent fine structure and test for hot spot motion.

\subsection*{PSR J1810+1744}

Our fit follows the assumptions of \citet{romani2021psr}, except that we allow for possible errors in the photometric zero points. This results in small changes in the fit parameters and a substantial increase in the distance and $L_H$ uncertainties. The neutron star mass is decreased from that paper's value by $\sim 0.5 \sigma$.

\subsection*{PSR J1959+2048}

This, the original BW pulsar, is of special interest, since \cite{van2011evidence}, with an approximate treatment of DH effects, infer that it may have a pulsar mass as large as $2.4\pm0.12M_\odot$. Recently, evidence of an eclipse of the pulsed $\gamma-$ray emission from J1959 has been presented in \citep{Clark:2021}. With a small $\sim 0.1R_\odot$ companion, this requires a binary inclination $i > 83^\circ$, far from the result of optical LC modeling in \cite{van2011evidence}. Some support for this edge-on view also comes from evidence for an X-ray eclipse in \citet{kandel2021xmm}.

With our improved treatment of gravity darkening, a simple DH model gives  $i= 65.7^\circ\pm 2.4^\circ$ and $\chi^2/$DoF = $138/89$. However, looking at the light curves, one can see a small phase shift in the maximum, with the peak somewhat broadened and excess flux at phases $\phi\sim 0.75–0.9$, especially in the bluer bands. In \cite{kandel2020atmospheric}, we showed how this excess can be explained by both a Gaussian hotspot and a banded wind. Because the LC maxima are relatively flat, the best-fit $i$ is $\sim 64^\circ$. Strong gravity darkening might also produce such a flat maximum, but the low front-side temperature of J1959 ($\lesssim 5000\,$K) together with a relatively low fill-factor of $\sim 0.8$ precludes this possibility.

For our analysis, we impose a prior on $i$ of $(83^\circ, 90^\circ)$, where the bounds are the result of geometrical constraints imposed by the observation of $\gamma-$ray eclipse. At such a high inclination, the lightcurve becomes relatively narrow at any reasonable value of the fill factor. Thus, to match the observational data, we require a spread of heat away from the maximum. While a banded wind can achieve this, a model with heat diffusion fits the data the best, resulting in $i=85.2^{+2.9}_{-1.2}$, with $\chi^2/$DoF=$128/88$. The angular scale of diffusion is $17.1^\circ \pm 4.2^\circ$. While this model fits data reasonably well, it is statistically worse than a model with an inclination as a free parameter. Fig \ref{fig:4lcb} shows the light curves for the best-fit model with inclination bound by $\gamma$-ray eclipse (first cycle) and one with free inclination (second cycle). $V$ and $R$ fit residuals for the two models are shown in the lower panels. This substantial tension between the gamma-ray eclipse and the optical light curve modeling is worrisome, and shows that this binary should be re-measured with modern multi-band photometry, to see if systematics affect the limited existing data or if additional physical effects are required to obtain a good fit.

\subsection*{PSR J2051$-$0827} Recently \citet{2022arXiv220809249D} have described simultaneous multi-color observations and ICARUS fitting of this BW. We do not attempt to re-fit here, but for comparison report the parameters of that study. They find significance evidence for a hot spot before 2011 (parameters not given), but that this spot is no longer prominent in 2021.

\subsection*{PSR J2052+1219}
All model fits require a fill factor $\sim 1$, so we fix $f=0.99$. Lacking RV data, we fix $M_{\rm NS}=1.5\,M_\odot$. A DH model gives $\chi^2/$DoF of $165/81$, with best-fit $i=59.8^\circ\pm 2.2^\circ$. Adding a hotspot reduces the $\chi^2$/DoF to $145/76$, with $i=60.6^\circ\pm 2.0^\circ$. However, the hotspot is rather large with a radius $\sim 39^\circ$ and a temperature-excess of $\sim 200\%$ of the base temperature. A WH model gives $\chi^2$/DoF to $153/80$, with a very similar inclination estimate of $i=59.3^\circ\pm 2.0^\circ$. After penalizing for the extra DoF in the HS model, the WH model is preferred at a marginal level of 60\% over the HS model. Physically, very shallow gradients (large effective spot radius) seem more natural for wind flow than a magnetic pole, so we prefer WH on these grounds. With very similar $i$, either model appears to capture the overall heating pattern. With good overall light curve matches (Fig \ref{fig:4lcb}), we infer that the large $\chi^2/$ DoF is the result of low-level stochastic flaring or possibly under-estimation of photometry errors.
With this inclination, the companion mass is  $M_c \sim 0.041 (M_{\rm NS}/1.5\,M_\odot)^{2/3}$, typical for a BW.

\begin{deluxetable*}{lcccc}
\tabletypesize{\footnotesize}
\tablewidth{0pt}
\tablecaption{Mass estimates of some BWs and RBs for which RV measurements are available.\label{tab:final_mass}}
\tablehead{
\colhead{Name} & \colhead{$i\, (\mathrm{ deg})$} & \colhead{$K_{\rm CoM}\, (\mathrm{ km/s})$} & \colhead{$M_{\rm PSR}\,(M_\odot)$} & \colhead{$M_{\rm C}\,(M_\odot)$}}
\startdata
J$0952-0607$&$59.8^{+2.0}_{-1.9}$ & $376.1\pm 5.1$&$2.35\pm 0.17$&$0.032\pm 0.002$\\
J$1301+0833$&$46.6^{+2.5}_{-2.2}$ &$274.8\pm 8.1$&$1.60_{-0.25}^{+0.22} $ & $0.036\pm 0.006$\\
J$1311-3430$& $68.7^{+2.1}_{-2.0}$ & $641.2\pm 3.6$&$2.22\pm 0.10$ &$0.012\pm0.0006$\\
J$1653-0158$& $72.8^{+4.0}_{-4.0}$ &$700.2\pm 8.0$ &$2.15^{+0.16}_{-0.16}$ &$0.014\pm 0.001$ \\
J$1810+1744$&$66.3^{+0.5}_{-0.5}$ & $462.9\pm 2.2$& $2.11\pm 0.04$& $0.064\pm 0.001$ \\ %https://www.overleaf.com/project/6275c5b89e22a52a919a491e
J$1959+2048$&$85.3^{+2.3}_{-1.2}$ & $334.6\pm 3.6$& $1.55_{-0.05}^{+0.06}$& $0.024\pm 0.001$\\
\enddata
 \label{table:final_fit}
\end{deluxetable*}

\section{Discussion and Summary}
\subsection{Thermal emission from the companion}
We find that a basic direct (photon) heating model is insufficient to represent the light curves of many BWs and RBs. There is statistically significant improvement in most model fits when adding a localized hot spot and/or global winds, and improved treatments of gravity and limb darkening. We compute multiple models for each object, including HS, WH, and diffusion, before deciding on the best, using the AIC. In the cases where the AIC did not select a preferred model and the competing models fit the light curve well, both give very similar binary kinematic parameters. In other words a well-matched light curve gives a reliable mass, even when the fit model is not unique.

However, variability can be an important factor in this modeling. The hot spot phases and brightness can vary, especially for RB, and may even show secular trends \citep[e.g][]{van2016active}. When correctly modeled, the underlying heating pattern, and the fit binary kinematic parameters should be stable. Other variability issues arise from the optical/X-ray flaring activity seen in strongly heated spiders. Such flares are best isolated in simultaneous multi-color photometric observations. It is important to excise these events before forming the `quiescent' light curve needed for the steady heating model and fit of the orbital parameters. An important, but often subtle effect is the presence of a non-thermal veiling flux. If associated with the IBS this may be phase dependent and may also exhibit secular variations. When bright enough to dominate at binary minimum (as for J1653), this too can be an important complication in fitting the quiescent heating pattern.

We find that $\sim$ half of the BW sources in our study prefer models with a hotspot. In Table \ref{tab:heating} we separate the companions' integrated thermal surface emission into day side $L_{\rm comp}$ and hotspot $L_{\rm hs}$ components. The hotpots often appear far from the direct heating maximum at the sub-pulsar point, and so can represent substantial changes in the local surface flux (and corresponding light curve features) with only modest energy input. The largest $L_{\rm hs}/L_{\rm comp}$ ratio is $\sim 0.25$ for J1653. Such hotspots are even more common in the RBs where the heavier ($\ge 0.08\,M_\odot$) companions should have core fusion and strongly convective envelopes. Since tidal locking ensures rapid rotation, the RBs should support strong $\alpha-\Omega$ dynamos resulting in large (but constantly refreshing) dipole magnetic fields. Such fields can direct IBS-generated particles to the magnetic poles \citep{2017ApJ...845...42S}, resulting in substantial hotspots, whose positions may vary with epoch. Although the BW companions lack core fusion, the very large front-back temperature gradients may well drive internal convection, resulting in similar dynamo-supported global fields. Indeed the objects preferring hotspots tend to have stronger heating. A recent two-epoch photometric study of BW J2051$-$0827 \citep{2022arXiv220809249D} suggests that hot spot variability may occur in BWs, as well.

We conclude that direct heating dominates energetically. Since pulsar SEDs are dominated by the GeV $\gamma$-rays we naturally assume that $\gamma$-rays will typically dominate the companion heating. But it is important to recall that the {\it Fermi}-observed $\gamma$-ray flux is from a single slice through the $\gamma$-ray beam along the Earth line-of-sight (for pulsars with aligned spin and orbital angular momenta this will be at co-latitude $i$). This cut may not represent the $\gamma$-ray flux intercepted near the orbital plane by the companion; for spin aligned pulsars this is the spin equator. Most modern outer magnetosphere models have the $\gamma$-ray flux concentrated to this equatorial plane, so in general we expect the companion-intercepted heating will be larger than that seen by {\it Fermi}, although the correction factors are highly model dependent \citep{draghis2019multiband}.

For most MSP, the observed photon (i.e.\,$\gamma$-ray) flux represents only a modest fraction of the spin-down power; the remainder is assumed to be carried off as the $e^\pm / B$ pulsar wind. For the BWs in our sample if we (naively, incorrectly) assume that the $\gamma$-ray flux is isotropic, we find that it represents $<0.25 I \Omega {\dot \Omega} = {\dot E}/4$. Table \ref{tab:heating} lists this isotropic $\gamma$-ray efficiency $f_{iso}$ for a standard moment of inertia $I=10^{45}{\rm g\,cm^2}$ (i.e. $I_{45}=1$). The very low $\gamma$-ray flux of J0636 indicates that our small $i$ view is outside of the main $\gamma$-ray beam. Three sources have a large isotropic $\gamma$-ray efficiency: For J2052, our photometric $d$ is substantially larger than the Table 4 $DM$ distance values and may be an overestimate; at the 2.4\,kpc DM distance it would have a more typical $f_{iso}=0.13$. From our fitting below, J1810 has a large pulsar mass (Table 3), so that we expect $I_{45}=2-3$, reducing $f_{iso}$ to 0.19-0.28. For J1311, the very large $f_{iso}=1.4$ may well be produced by a combination of the two factors; the inferred NS mass is large and $d$ exceeds the $DM$ estimates. Here an independent (eg. optical) parallax distance would be quite valuable. Of course all these large values should be mitigated by $\gamma$-ray beaming preferentially to the Earth line-of-sight.

Since in modern models the pulsar wind power (as well as the $\gamma$-ray pulsed emission) is concentrated to the spin equator, we approximate this in our direct heating models by distributing the heating power as $\propto {\rm sin}^2 \theta$, with $\theta$ measured from the spin axis. With this distribution we can write the model-fit heating flux's value for the Earth line-of-sight $f_H$. It now becomes interesting to examine $\eta_\gamma = f_H/f_\gamma$. If we assume that the direct heating is $\gamma$-ray pulsed flux $\eta_\gamma > 1$ tells us that there is more $\gamma$-emission heating the companion near the spin equator than at the Earth sightline, and vice-versa. Of course, we can also compare the integral heating flux required with ${\dot E}$, via $\eta_{\dot E} = L_H/{\dot E}$. At $\eta_{\dot E} = 1$, ${\rm sin}^2\theta$ heating would require the full $I_{45}=1$ spindown power. For $\eta_{\dot E}>1$ we infer some combination of spindown power more equatorially focused, large $I_{45}$ and smaller $d$. J1311 certainly requires some or all of these effects to reconcile the observed heating with the pulsar energy loss rate.

\begin{deluxetable*}{lccccc}
\tabletypesize{\footnotesize}
\tablewidth{0pt}
\tablecaption{BW distances and Shklovskii-corrected magnetic fields\label{tab:dist_summary}}
\tablehead{\colhead{Name} & \colhead{$d$ (kpc)} & \colhead{CL02(kpc)} & \colhead{Y17(kpc)} &\colhead{Parallax(kpc)}&\colhead{$B_{\rm int}(10^8$G)}}
\startdata
J$0023+0923$& $0.59^{+1.39}_{-0.39}$ & 0.69& 1.25&$-$&$1.82^{+0.05}_{-0.17}$ \cr
J$0636+5128$& $0.63^{+1.44}_{-0.43}$ & 0.49 & 0.21 &$1.1^{+0.6}_{-0.3}$\tablenotemark{a}&$1.0^{+0.01}_{-0.06}$\cr%DM=11.1 Timing Pi Arzoumanian 18
J$0952-0607$ & $6.26^{+0.36}_{-0.40}$ & 0.97& 1.74&$-$&$0.61\pm0.02$\cr
J$1301+0833$& $1.77^{+0.10}_{-0.11}$&  0.67 &1.23&$-$&$0.96\pm0.03$\cr
J$1311-3430$& $3.01\pm 0.15$& 1.4 & 2.43&$-$&$2.29\pm0.03$\cr
J$1653-0158$& $0.87^{+0.07}_{-0.08}$&  & &$0.57^{+0.44}_{-0.18}$\tablenotemark{b}&$0.36\pm0.04$\cr%GAIA DR3 1.77+/-0.78 --> 0.57+0.44-0.18  was $0.53^{+0.63}_{-0.19}$
J$1810+1744$& $2.86\pm 0.13$& 2.00& 2.36 &$1.54^{+7.61}_{-0.70} $\tablenotemark{b}&$0.77\pm 0.01$\cr    %GAIA DR3 0.65+/-0.54 --> 1.54+7.6-0.70
J$1959+2048$& $2.25\pm 0.11$ & 2.49& 1.73&$2.57^{+1.84}_{-0.77}$\tablenotemark{c}& $1.19\pm0.01$\cr %This is VLBI, Romani et al 2022
%                    also GAIA DR3 1.19+/-1.36  --> 0.84+inf-0.45
J$2052+1219$& $5.57^{+0.47}_{-0.55}$& 2.4 & 3.92&$-$&$0.26\pm 0.22$\cr
\enddata
\tablenotetext{\tiny a}{Timing parallax \citet{arzoumanian2018nanograv};\,$^b$GAIA DR3;\,$^c$VLBI parallax \citet{2022ApJ...930..101R}}
 \label{table:objects_dist}
\end{deluxetable*}

%done updating
\begin{deluxetable*}{lccccccc}
\tabletypesize{\footnotesize}
\tablewidth{0pt}
\tablecaption{BW Derived and Observed Heating\label{tab:heating}}
\tablehead{\colhead{Name} & \colhead{$f_\gamma$} & $\eta_{iso}$ & \colhead{$f_{\rm H}$\tablenotemark{a}} &  \colhead{$\eta_{\dot E}$} & \colhead{$\eta_\gamma$} & \colhead{$L_{\rm comp}$\tablenotemark{b}} & \colhead{$L_{\rm hs}$\tablenotemark{b}}\\
 & $10^{-12}$erg/s/cm$^2$ & & $10^{-12}$erg/s/cm$^2$ & & &  $10^{30}$erg/s & $10^{30}$erg/s}
\startdata
J$0023+0923$& $7.3\pm 1.4$ &0.019 & $40.6^{+430}_{-38.3}$ & 0.076 & 5.6 &  $8.0\pm 1.0$ & $1.5^{+0.7}_{-0.5}$\cr
J$0636+5128$ & $<0.5$& $<0.004$ &$18.9^{+230}_{-17.7}$ & 0.33 & $>$38& $4.1^{+0.8}_{-0.7}$ & $0.1^{+0.04}_{-0.02}$\cr
J$0952-0607$ & $2.2\pm 0.5$& 0.154 &$0.91^{+0.23}_{-0.22}$& 0.057 & 0.41 &$108^{+10}_{-11}$&$-$\cr
J$1301+0833$& $10.6\pm 1.5$& 0.060 & $10.8^{+3.9}_{-2.84}$ & 0.071 & 1.0 &$16.6^{+1.7}_{-1.8}$&$-$\cr
J$1311-3430$& $64.7\pm 1.9$& 1.40 & $273^{+106}_{-85.6}$ & 4.56 & 4.2 &  $321^{+80}_{-62}$ & $16^{+19}_{-7}$\cr
J$1653-0158$& $33.7\pm 1.8$& 0.26 & $65.0^{+26.3}_{-18.1}$ &0.11 & 1.9 &  $7.3\pm 0.9$ & $1.9^{+0.9}_{-0.6}$\cr
J$1810+1744$& $22.4\pm 1.3$&0.56 & $47.8^{+5.9}_{-5.6}$ & 1.05 & 2.1 &  $187^{+17}_{-19}$& $19\pm 2$\cr
J$1959+2048$& $17.9\pm 1.9$&0.068 & $105\pm 9.5$& 0.27 & 5.9 &$110^{+7}_{-8}$ & $-$ \cr
J$2052+1219$& $6.5\pm 1.0$&0.720 & $3.5^{+2.3}_{-1.5}$ & 0.35 & 0.54 &$72.1^{+13.8}_{-12.0}$&$-$
\enddata
\tablenotetext{\tiny a}{$f_H=3 {\rm sin^2}i \,L_h/(8\pi d^2)$, i.e. assuming $\sin^2\theta$ $\gamma$-ray beaming, orbital and spin momenta aligned.}
\tablenotetext{\tiny b}{Bolometric luminosity of the hotspot and the companion star without hotspot. Uncertainties generated through MCMC chain of the photometric fitting.}

\end{deluxetable*}

\subsection{Mass distributions}
\begin{figure}
    \centering
    \includegraphics[scale=0.35]{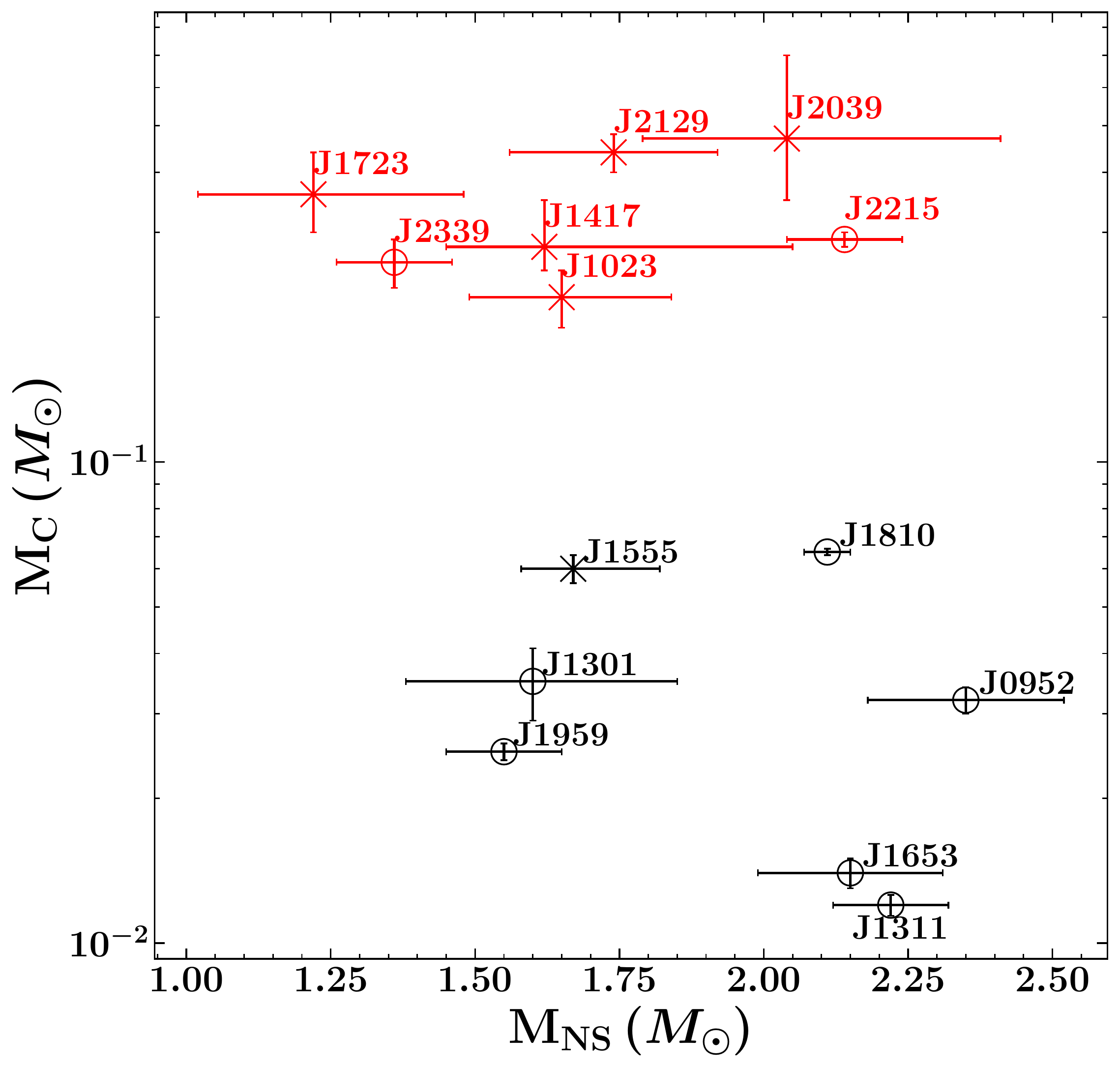}
    \caption{The companion mass versus NS mass for all BWs (black points) and RBs (red points) for which a reliable mass measurement exists. Masses for objects with circled points come from this paper; cross points are RBs from \cite{strader2019optical} and BW J1555 from \citet{2022MNRAS.512.3001K}. }
    \label{fig:mass_distri}
\end{figure}

Figure \ref{fig:mass_distri} shows the distribution of NS and companion masses for both BWs and RBs, with most reliable measurements of BWs coming from this paper. The NS masses of RBs have a wide spread and are on average lower than those of BWs. For companion masses, the BW/RB separation remains clear, with BWs having a low companion mass $\sim 0.01- 0.06\,M_\odot$, and RBs having a higher companion mass $\sim 0.2-0.5\,M_\odot$. However, the detection of relatively heavy BW companions in J1810 and J1555 suggests that the mass gap between these two populations is a factor $\sim 3\times$.

In \cite{chen2013formation}, it was argued that RBs and BWs are distinct evolutionary populations, with irradiation efficiency determining the formation channel; in this picture RBs do not evolve into BWs. On the other hand, numerical simulations in \cite{benvenuto2014understanding} suggests that $P_B<1$\,d RBs will indeed evolve into BWs. Our sample does not distinguish these two scenarios, although a decrease in the BW-RB companion mass gap could be seen as supporting an evolutionary connection.

\begin{deluxetable}{lccl}
\tabletypesize{\footnotesize}
\tablewidth{0pt}
\tablecaption{BW Companion Mean Densities\label{table:objects_size}}
\tablehead{\colhead{Name} & \colhead{$P_b$(h)} &\colhead{$\rho$ (g/cm$^3$)} & \colhead{$R/R_{\rm Y_e=Y_\odot}$} }
\startdata
J$0023+0923$ & 3.33& 22.8$\pm 0.8$& 0.92\cr
J$0636+5128$ & 1.60 & 21.3$\pm 0.3$& 1.77\tablenotemark{a}\cr
J$0952-0607$ & 6.42 & 40.1$\pm 2.5$& 1.12 \cr
J$1301+0833$& 6.48 & 36.1$\pm 7.5$& 1.25 \cr
J$1311-3430$& 1.56 & 28.6$\pm 1.4$& 1.52\tablenotemark{a}\cr
J$1653-0158$& 1.25 & 33.3$\pm 2.5$ & 1.61\tablenotemark{a}\cr
J$1810+1744$& 3.56 & 30.8$\pm 0.5$ & 1.94 \cr
J$1959+2048$& 9.17 & 28.6$\pm 1.9$ & 1.04\cr
J$2051-0827$& 2.37 & 20.2$\pm 0.6$ & 1.60\tablenotemark{b} \cr %Dhillon M=0.039+/-0.011 f1=0.88+/-0.02
J$2052+1219$ & 2.75 & 33.3 &1.64 \cr
\enddata

\tablenotetext{\tiny a}{UCXRB progenitor; cold radius for $Y_e=0.5$.}
\tablenotetext{\tiny b}{\citet{2022arXiv220809249D}}
 \label{table:objects_size}
\end{deluxetable}

In Table \ref{table:objects_size}, we list the inferred mean density of our companions. With our fit parameters, these are surprisingly similar, varying by less than $2\times$. We also list the ratio of their volume equivalent radius to that of a cold degenerate object of the companion mass. The $T=0$ radius for a $\Gamma=5/3$ degenerate object with $Y_e$ electrons per nucleon is $R_{\rm cold}=0.0126\left( \frac{M_C} {M_{\odot}}\right)^{-1/3}(2Y_e)^{5/3}R_\odot$. In most BW evolutionary scenarios, mass transfer and subsequent companion evaporation are initiated well before core exhaustion. In this case we might expect a $\sim$ solar metalicity for the companion with $Y_e\approx 0.833$. We see that most companions are inflated above this value (up to nearly $2\times$ for the strongly heated J1810). A few objects have small $R/R_{Y_\odot}$, (e.g. J0023); for these we might assume some core enrichment before evaporation commences, giving smaller cold $R$ and stronger inflation. On the other hand the shortest period BWs are believed to be the descendants of ultra-compact X-ray binaries, with hydrogen depleted secondaries, so for the three objects with $P_b<2$\,h we can assume $Y_e=0.5$. This is supported by spectroscopy of J1311 and J1653, where the absence of H features shows that even the photosphere is H depleted \citep{romani20142fgl}.

\begin{figure}[t!]
\centering
\hspace*{-5mm}\includegraphics[scale=0.45]{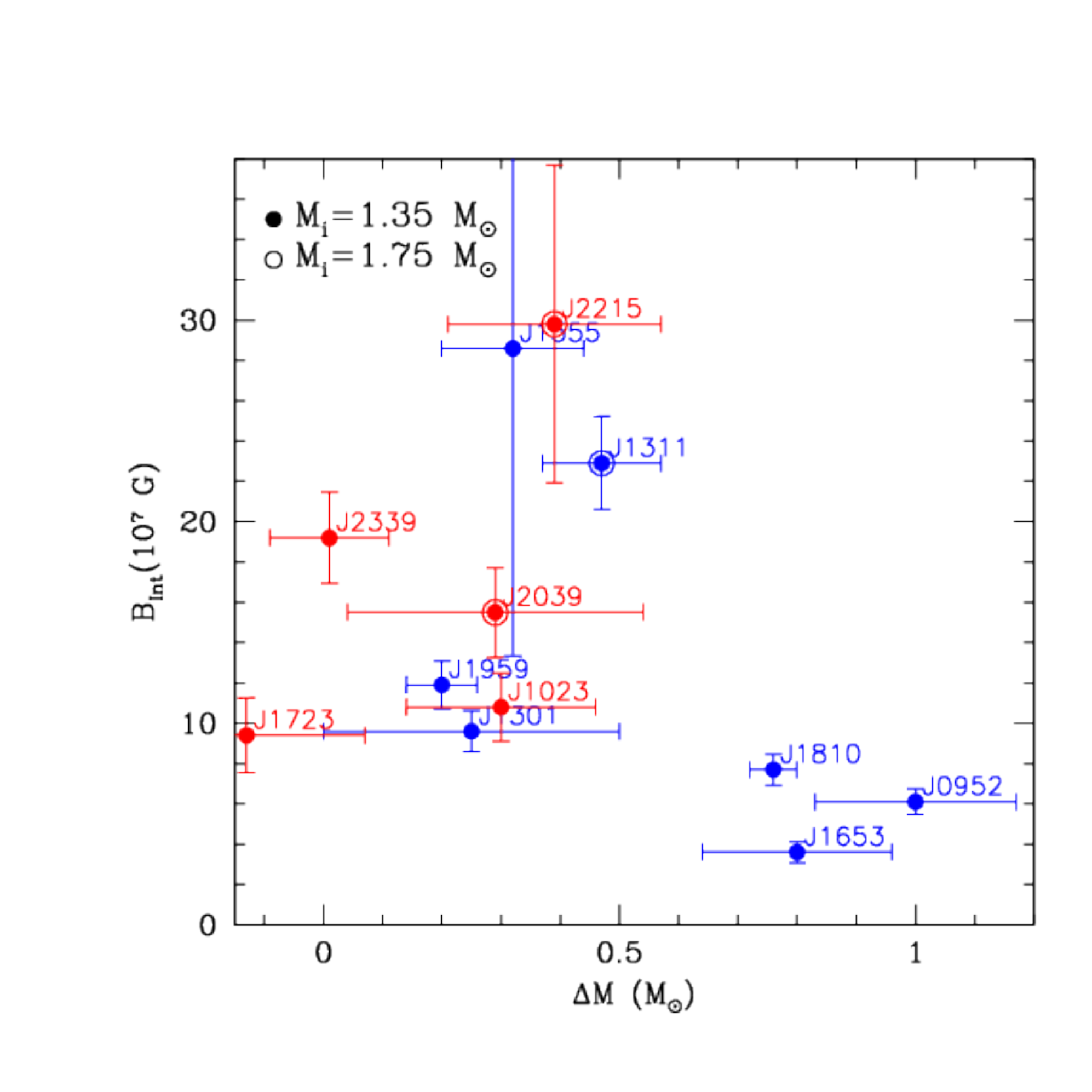}
\caption{Shklovskii-corrected surface dipole magnetic field vs. accreted mass for some known BWs (blue) and RBs (red), assuming photometric distances except when more accurate parallax distances are available. Note that for three PSRs (J1311, J2039 and J2215) we assume that they were born as higher mass neutron stars. While the RBs show no trend the BWs seem to have lower magnetic fields at higher accreted (and final) mass.}
\label{fig:B_field}
\end{figure}

\subsection{Relationship between NS mass and B field}

Binary pulsars have lower inferred dipole fields than isolated neutron stars, and it has long been proposed \citep[e.g.,][]{bisnovatyi1974magnetic, romani1990unified} that past accretion has decreased young pulsar magnetic fields to millisecond-pulsar values. It is unclear whether the amount of mass accreted (as in simple burial models) or the duration of the accretion phase (as in heating-driven ohmic decay) is most important \citep[see][for a review]{mukherjee2017revisiting}. Note that small MSP $P_s$ are not, of themselves, precise probes of accretion history. Since the angular momentum needed for spin-up to breakup periods is only $\sim 0.1\,M_\odot$ and since birth masses seem to vary by at least this amount it is unlikely that even high precision mass measurements can determine the mass gain with sufficient accuracy to probe angular moment addition (although statistical studies might prove useful). Instead $P_s$ tracks the neutron star dipole field in equilibrium spin-up models, thus telling us something about the accretion rate at the end of spin-up, and subsequent spindown. The best hope for a probe of dipole field reduction occurs if large amounts of mass are needed to decrease the field from initial Terra-Gauss values to $\sim 10^8$G levels required for small $P_s$. Even then the situation is complicated, since MSP population studies suggest that the {\it initial} masses $M_i$ of these pulsars are bimodal with a dominant component of $M_i \le 1.4M_\odot$ and a 20\% sub-population with $M_i\sim 1.8 M_\odot$ \citep[][and references therein]{2016arXiv160501665A}.

We make a first attempt at such comparison in Figure \ref{fig:B_field} where the Shklovskii-corrected intrinsic dipole surface field $B_{\rm int}$ for an assumed moment of inertia $I_{45}=1$ is plotted against inferred mass increase. Table \ref{table:objects_dist} $B_{\rm int}$ values are computed assuming the nominal photometric distance uncertainty; for the plot we add an 10\% additional uncertainty in quadrature to acknowledge $I_{45}$ variation with the uncertain mass. Three BWs that have particularly high accreted mass (J0952 $\Delta M \approx 1.0 M_\odot$, J1653$-$0158 $\Delta M \approx 0.8M_\odot$, and J1810+1740 $\Delta M \approx 0.76 M_\odot$, assuming a start at typical $1.35\, M_\odot$) also have some of the lowest $B_{\rm int}$ known. However, J1723 has a relatively low surface field and has apparently accreted very little mass (although its mass measurement has large error bars, and this object has not been subject to the uniform fits of our study). Conversely the well-measured J1311, J2039 and J2215 are heavy neutron stars. Their $B_{\rm int}$ fields also seem substantial at $>1.5\times 10^8$\,G, although with the possible photometric distance errors above, the Shklovskii corrections remain quite uncertain. A possible solution is that these systems may be from the 20\% subset of MSP starting the accretion phase from $\sim 1.8\,M_\odot$. This is assumed in Figure \ref{fig:B_field}. At this point we can only conclude that the largest $M$, smallest $B_{\rm int}$ and shortest $P_s$ appear associated with substantial mass increase -- it is likely that other factors are important in determining $B_{\rm int}$ and we will need more, and better ($B_{\rm int}$, $\Delta M$) pairs to draw detailed conclusions.

\begin{figure}[t!]
\centering
\vspace*{-15mm}
\hspace*{-5mm}
\includegraphics[scale=0.44]{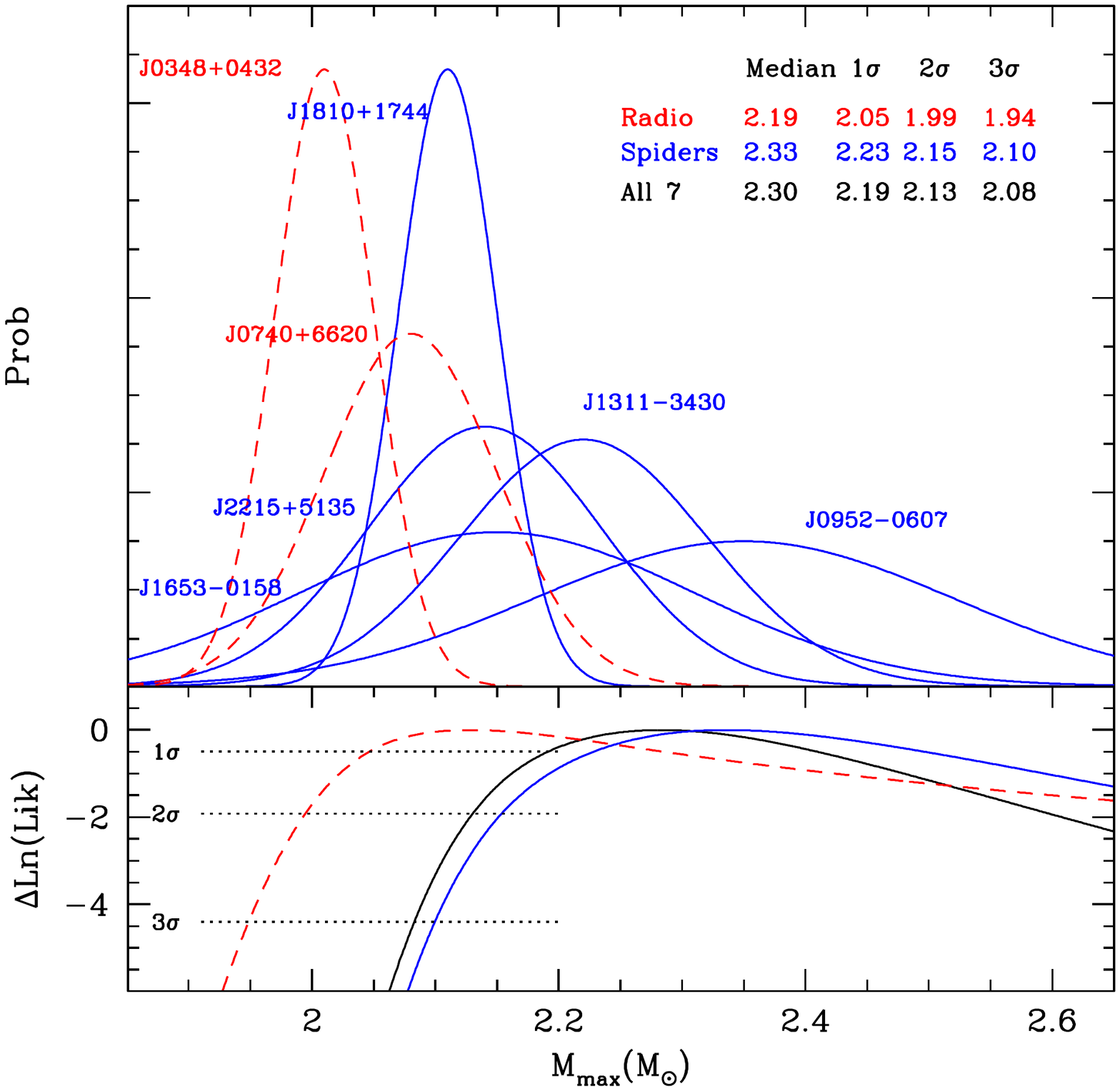}
\vspace*{-28mm}
\caption{Normalized probability distributions of mass estimates for MSPs with $M_{\rm NS} > 2.0\,M_\odot$. The dashed curves show the mass estimates for the two most massive radio-selected pulsars with white dwarf (WD) companions, measured from pulse timing (supplemented by WD atmosphere modeling for J0348). The solid curves show spider binary mass estimates, relying on companion spectrophotometry. The bottom panel shows the cumulative probability distributions for $M<M_{\rm max}$, for the radio objects, the spiders, and all six pulsars. Lower bounds on $M_{\rm max}$ ($1\sigma$, $2\sigma$, and $3\sigma$) for these distributions are listed in the legend at top.}
\label{fig:masses}
\end{figure}

\subsection{Mass implications on NS EoS}

Shapiro delay measurements of radio pulsar PSR J0348+0432 give $2.01\pm0.04 \, M_\odot$ \citep{antoniadis2013massive} and PSR J0740+6620 give $2.08\pm 0.07\, M_\odot$ \citep{fonseca2021refined}. However, the quoted errors are at the $1 \sigma$ level, and so these data do not require $M_{\rm max}> 2\,M_\odot$ at high statistical confidence. Since the impact on EoS modeling is driven by the minimum required $M_{\max}$, higher masses, with tighter estimates should provide even greater physics impact. The spiders pulsars are good candidates for such measurements and, indeed, J1810 is the first individual object for which a $\sim 3\sigma$ lower bound on the mass exceeds $2\,M_\odot$ \citep{romani2021psr}. We have argued above that our improved, uniform treatment of BW LC and spectroscopy data results in more stable and better-determined mass measurements, relatively immune to modeling uncertainty. Our task now is to combine these into a global constrain on $M_{\max}$. Figure \ref{fig:masses} shows the mass uncertainty ranges for the objects in this paper with central values $> 2M_\odot$.

Several approaches can be used to estimate $M_{\rm max}$. One option is to model the full distribution of (binary) NS masses and see if an upper cutoff is required; \cite{alsing2018evidence}, for example, determine that $M_{\rm max}$ is within a $1\sigma$ range of 2.0--2.2\,$M_\odot$. Here we only attempt to determine a lower bound to $M_{\rm max}$. Assuming that our heavy NS sample is drawn from a uniform population of lower bound $M_1\equiv 1.8\,M_\odot$ and upper bound  $M_{\rm max}$,  $U(M_1, M_{\rm max})$, the marginal distribution of individual observations $\{M_i,\sigma_i\}$ is:
\begin{align}
    p(M_i)&=\int_0^{\infty}{\rm d}\mu_i\,N(M_i|\mu_i;\sigma_i)\,U(\mu_i|M_{1},M_{\rm max})\\
    &=\frac{{\rm Erf}\left[\frac{M_i-M_{1}}{\sqrt{2}\sigma_i}\right]-{\rm Erf}\left[\frac{M_i-M_{\rm max}}{\sqrt{2}\sigma_i}\right]}{2(M_{\rm max}-M_1)}
\end{align}
For $n$ observations, the log(likelihood) is
\begin{align}
    \log \mathcal{L}&=-n\log(M_{\rm max}-M_{1})\nonumber \\
    &+\sum_{i=1}^n\log\left({\rm Erf}\left[\frac{M_i-M_{1}}{\sqrt{2}\sigma_i}\right]-{\rm Erf}\left[\frac{M_i-M_{\rm max}}{\sqrt{2}\sigma_i}\right]\right)
\end{align}
The log(likelihood) curves for different sample sets (Radio and Spiders) are plotted in the lower panel of Fig. \ref{fig:masses}
and the median value, as
well as the $1-, 2-$ and $3-\sigma$ lower bounds on $M_{\rm max}$ are listed for each sample. At the $1\sigma$ level, $M_{\rm max}$ inferred from spiders is higher than that inferred from radio  measurements by $0.18\,M_\odot$, suggesting that NSs in the spider systems are significantly heavier that those in NS-white dwarf binaries. In our uniform re-fit of these spiders we have marginalized over the photometric parameters. We have also shown that with good fits to high quality LC data, we can select between various physically-motivated heating models. When different models' fit-quality is similar, we find that the orbital parameters determining the fit mass are similar too. Thus we have reduced systematic uncertainties, and it becomes worth discussing statistical bounds at higher confidence levels. We can now say with high ($3\sigma$ one-sided lower bound) statistical confidence that $M_{\rm max} > 2.08\,M_\odot$, and that at $\sim 1\sigma$ significance $M_{\rm max}> 2.19\,M_\odot$ is preferred.

\begin{figure}[ht!!]
    \centering
    \includegraphics[scale=0.35]{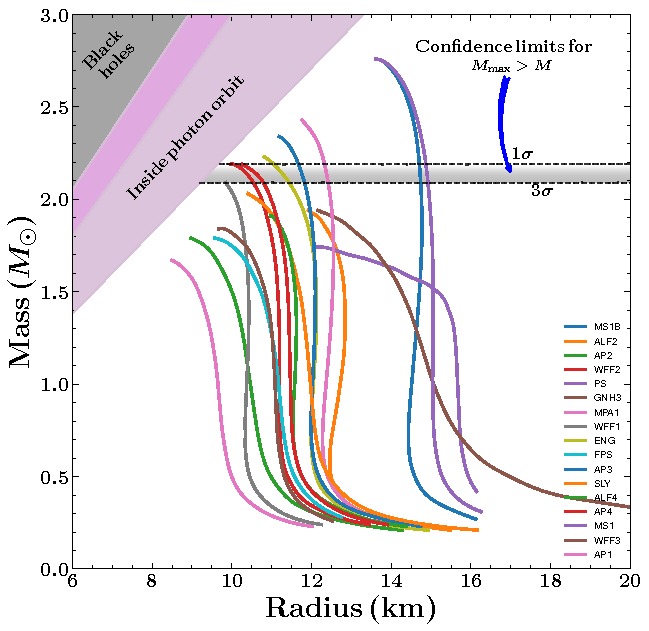}
    \caption{Mass-radius curves for a range of EoS. Figure adapted from \citet{ozel2016masses}; see that paper for the EoS labels. Note that any viable EOS must rise above the $M_{\rm max}$ lower bound. As this increases many otherwise viable EoS are excluded at high confidence level.}
    \label{fig:EoS}
\end{figure}

For each EoS, integration of the Oppenheimer-Volkoff equations gives a different $M-R$ curve. Curves for various literature EoS are shown in Figure \ref{fig:EoS}. Although some of these are already deprecated by radius and tidal deformation measurements of common $\sim 1.4\,M_\odot$ NS, if one is interested in the physics of extreme conditions, the most important aspect of these curves is probed by the roll-off at their maximum. This is because only high mass objects achieve the core densities to probe such physics. `Soft', compressible EoS are already strongly excluded by the $\sim 2 M_\odot$ radio MSP. But the compressibility at $\rho > 4 \rho_{\rm sat}$ is best probed at higher masses. In general EOS with non-nucleonic phases, such as hyperons or condensates cannot reach these masses. Conversely, to reach masses far above $2M_\odot$ a phase transition from nucleonic to even less compressible matter might be required. In any event, the bounds from figure \ref{fig:masses} are the tightest constraints in this region. We have transferred the 1$\sigma$ and $3\sigma$ limits from all spiders to Figure \ref{fig:EoS}. Note that only five of the listed models (two with implausibly large radii at $\sim 1.4M_\odot$) survive the $1\sigma$ bound. Extending to $3 \sigma$, only two more survive. It is clear that astrophysics community's efforts to pin down neutron stars in the $M-R$ plane, substantially enhanced by our spider measurements, are dramatically narrowing the options for the dense matter EoS.

\bigskip

We thank Alex Filippenko and colleagues for collaboration on many of the observations re-analyzed in this paper. We also thank Rene Breton for useful discussions on spider light curve fitting. DK and RWR were supported in part by NASA grants 80NSSC17K0024390 and 80NSSC17K0502.

\bibliographystyle{aasjournal}
\bibliography{mainpaper}

\begin{thebibliography}{}
\expandafter\ifx\csname natexlab\endcsname\relax\def\natexlab#1{#1}\fi
\providecommand{\url}[1]{\href{#1}{#1}}
\providecommand{\dodoi}[1]{doi:~\href{http://doi.org/#1}{\nolinkurl{#1}}}
\providecommand{\doeprint}[1]{\href{http://ascl.net/#1}{\nolinkurl{http://ascl.net/#1}}}
\providecommand{\doarXiv}[1]{\href{https://arxiv.org/abs/#1}{\nolinkurl{https://arxiv.org/abs/#1}}}

\bibitem[{{Akaike}(1974)}]{1974ITAC...19..716A}
{Akaike}, H. 1974, IEEE Transactions on Automatic Control, 19, 716

\bibitem[{Aldcroft {et~al.}(1992)Aldcroft, Romani, \&
  Cordes}]{aldcroft1992spectroscopy}
Aldcroft, T.~L., Romani, R.~W., \& Cordes, J.~M. 1992, The Astrophysical
  Journal, 400, 638

\bibitem[{Alsing {et~al.}(2018)Alsing, Silva, \& Berti}]{alsing2018evidence}
Alsing, J., Silva, H.~O., \& Berti, E. 2018, \mnras, 478, 1377

\bibitem[{Antoniadis(2021)}]{antoniadis2021gaia}
Antoniadis, J. 2021, Monthly Notices of the Royal Astronomical Society, 501,
  1116

\bibitem[{{Antoniadis} {et~al.}(2016){Antoniadis}, {Tauris}, {Ozel}, {Barr},
  {Champion}, \& {Freire}}]{2016arXiv160501665A}
{Antoniadis}, J., {Tauris}, T.~M., {Ozel}, F., {et~al.} 2016, arXiv e-prints,
  arXiv:1605.01665.
\newblock \doarXiv{1605.01665}

\bibitem[{Antoniadis {et~al.}(2013)Antoniadis, Freire, Wex, Tauris, Lynch, van
  Kerkwijk, Kramer, Bassa, Dhillon, Driebe, {et~al.}}]{antoniadis2013massive}
Antoniadis, J., Freire, P.~C., Wex, N., {et~al.} 2013, Science, 340

\bibitem[{Arzoumanian {et~al.}(2018)Arzoumanian, Brazier, Burke-Spolaor,
  Chamberlin, Chatterjee, Christy, Cordes, Cornish, Crawford, Cromartie,
  {et~al.}}]{arzoumanian2018nanograv}
Arzoumanian, Z., Brazier, A., Burke-Spolaor, S., {et~al.} 2018, The
  Astrophysical Journal Supplement Series, 235, 37

\bibitem[{Benvenuto {et~al.}(2014)Benvenuto, De~Vito, \&
  Horvath}]{benvenuto2014understanding}
Benvenuto, O.~G., De~Vito, M.~A., \& Horvath, J.~E. 2014, The Astrophysical
  Journal Letters, 786, L7

\bibitem[{Bisnovatyi-Kogan(1974)}]{bisnovatyi1974magnetic}
Bisnovatyi-Kogan, G. 1974, Soviet Astronomy, 18, 261

\bibitem[{Breton {et~al.}(2012)Breton, Rappaport, van Kerkwijk, \&
  Carter}]{breton2012koi}
Breton, R.~P., Rappaport, S.~A., van Kerkwijk, M.~H., \& Carter, J.~A. 2012,
  The Astrophysical Journal, 748, 115

\bibitem[{Callanan {et~al.}(1995)Callanan, Van~Paradijs, \&
  Rengelink}]{callanan1995orbital}
Callanan, P.~J., Van~Paradijs, J., \& Rengelink, R. 1995, The Astrophysical
  Journal, 439, 928

\bibitem[{Chen {et~al.}(2013)Chen, Chen, Tauris, \& Han}]{chen2013formation}
Chen, H.-L., Chen, X., Tauris, T.~M., \& Han, Z. 2013, The Astrophysical
  Journal, 775, 27

\bibitem[{Claret \& Bloemen(2011)}]{claret2011gravity}
Claret, A., \& Bloemen, S. 2011, Astronomy \& Astrophysics, 529, A75

\bibitem[{Clark {et~al.}(2021)Clark, M., \& Breton}]{Clark:2021}
Clark, C., M., K., \& Breton, R.~P. 2021, in {9th International Fermi
  Symposium}

\bibitem[{{Dhillon} {et~al.}(2022){Dhillon}, {Kennedy}, {Breton}, {Clark},
  {Mata S{\'a}nchez}, {Voisin}, {Breedt}, {Brown}, {Dyer}, {Green}, {Kerry},
  {Littlefair}, {Marsh}, {Parsons}, {Pelisoli}, {Sahman}, {Wild}, {van
  Kerkwijk}, \& {Stappers}}]{2022arXiv220809249D}
{Dhillon}, V.~S., {Kennedy}, M.~R., {Breton}, R.~P., {et~al.} 2022, arXiv
  e-prints, arXiv:2208.09249.
\newblock \doarXiv{2208.09249}

\bibitem[{Djorgovski \& Evans(1988)}]{djorgovski1988photometry}
Djorgovski, S., \& Evans, C.~R. 1988, The Astrophysical Journal, 335, L61

\bibitem[{Draghis \& Romani(2018)}]{draghis2018psr}
Draghis, P., \& Romani, R.~W. 2018, The Astrophysical Journal Letters, 862, L6

\bibitem[{Draghis {et~al.}(2019)Draghis, Romani, Filippenko, Brink, Zheng,
  Halpern, \& Camilo}]{draghis2019multiband}
Draghis, P., Romani, R.~W., Filippenko, A.~V., {et~al.} 2019, The Astrophysical
  Journal, 883, 108

\bibitem[{Fonseca {et~al.}(2021)Fonseca, Cromartie, Pennucci, Ray, Kirichenko,
  Ransom, Demorest, Stairs, Arzoumanian, Guillemot,
  {et~al.}}]{fonseca2021refined}
Fonseca, E., Cromartie, H., Pennucci, T.~T., {et~al.} 2021, The Astrophysical
  Journal Letters, 915, L12

\bibitem[{{Fruchter} {et~al.}(1988){Fruchter}, {Stinebring}, \&
  {Taylor}}]{1988Natur.333..237F}
{Fruchter}, A.~S., {Stinebring}, D.~R., \& {Taylor}, J.~H. 1988, \nat, 333,
  237, \dodoi{10.1038/333237a0}

\bibitem[{Green {et~al.}(2019)Green, Schlafly, Zucker, Speagle, \&
  Finkbeiner}]{green20193d}
Green, G.~M., Schlafly, E., Zucker, C., Speagle, J.~S., \& Finkbeiner, D. 2019,
  The Astrophysical Journal, 887, 93

\bibitem[{Husser {et~al.}(2013)Husser, Wende-von Berg, Dreizler, Homeier,
  Reiners, Barman, \& Hauschildt}]{husser2013new}
Husser, T.-O., Wende-von Berg, S., Dreizler, S., {et~al.} 2013, Astronomy \&
  Astrophysics, 553, A6

\bibitem[{Kandel \& Romani(2020)}]{kandel2020atmospheric}
Kandel, D., \& Romani, R.~W. 2020, The Astrophysical Journal, 892, 101

\bibitem[{Kandel {et~al.}(2021)Kandel, Romani, \& An}]{kandel2021xmm}
Kandel, D., Romani, R.~W., \& An, H. 2021, The Astrophysical Journal Letters,
  917, L13

\bibitem[{{Kandel} {et~al.}(2020){Kandel}, {Romani}, {Filippenko}, {Brink}, \&
  {Zheng}}]{2020ApJ...903...39K}
{Kandel}, D., {Romani}, R.~W., {Filippenko}, A.~V., {Brink}, T.~G., \& {Zheng},
  W. 2020, \apj, 903, 39, \dodoi{10.3847/1538-4357/abb6fd}

\bibitem[{{Kennedy} {et~al.}(2022){Kennedy}, {Breton}, {Clark}, {Mata
  S{\'a}nchez}, {Voisin}, {Dhillon}, {Halpern}, {Marsh}, {Nieder}, {Ray}, \&
  {van Kerkwijk}}]{2022MNRAS.512.3001K}
{Kennedy}, M.~R., {Breton}, R.~P., {Clark}, C.~J., {et~al.} 2022, \mnras, 512,
  3001, \dodoi{10.1093/mnras/stac379}

\bibitem[{{Kulkarni} {et~al.}(1988){Kulkarni}, {Djorgovski}, \&
  {Fruchter}}]{1988Natur.334..504K}
{Kulkarni}, S.~R., {Djorgovski}, S., \& {Fruchter}, A.~S. 1988, \nat, 334, 504,
  \dodoi{10.1038/334504a0}

\bibitem[{Mukherjee(2017)}]{mukherjee2017revisiting}
Mukherjee, D. 2017, Journal of Astrophysics and Astronomy, 38, 1

\bibitem[{Nieder {et~al.}(2020)Nieder, Clark, Kandel, Romani, Bassa, Allen,
  Ashok, Cognard, Fehrmann, Freire, {et~al.}}]{nieder2020discovery}
Nieder, L., Clark, C., Kandel, D., {et~al.} 2020, The Astrophysical journal
  letters, 902, L46

\bibitem[{{\"O}zel \& Freire(2016)}]{ozel2016masses}
{\"O}zel, F., \& Freire, P. 2016, Annual Review of Astronomy and Astrophysics,
  54, 401

\bibitem[{Ray {et~al.}(2012)Ray, Abdo, Parent, Bhattacharya, Bhattacharyya,
  Camilo, Cognard, Theureau, Ferrara, Harding, {et~al.}}]{ray2012radio}
Ray, P., Abdo, A., Parent, D., {et~al.} 2012, arXiv preprint arXiv:1205.3089

\bibitem[{Ray {et~al.}(2013)Ray, Ransom, Cheung, Giroletti, Cognard, Camilo,
  Bhattacharyya, Roy, Romani, Ferrara, {et~al.}}]{ray2013radio}
Ray, P., Ransom, S., Cheung, C., {et~al.} 2013, The Astrophysical journal
  letters, 763, L13

\bibitem[{Reynolds {et~al.}(2007)Reynolds, Callanan, Fruchter, Torres, Beer, \&
  Gibbons}]{reynolds2007light}
Reynolds, M.~T., Callanan, P.~J., Fruchter, A.~S., {et~al.} 2007, Monthly
  Notices of the Royal Astronomical Society, 379, 1117

\bibitem[{Romani(1990)}]{romani1990unified}
Romani, R.~W. 1990, Nature, 347, 741

\bibitem[{Romani {et~al.}(2014)Romani, Filippenko, \& Cenko}]{romani20142fgl}
Romani, R.~W., Filippenko, A.~V., \& Cenko, S.~B. 2014, The Astrophysical
  Journal Letters, 793, L20

\bibitem[{Romani {et~al.}(2015)Romani, Filippenko, \&
  Cenko}]{romani2015spectroscopic}
---. 2015, The Astrophysical Journal, 804, 115

\bibitem[{Romani {et~al.}(2012)Romani, Filippenko, Silverman, Cenko, Greiner,
  Rau, Elliott, \& Pletsch}]{romani2012psr}
Romani, R.~W., Filippenko, A.~V., Silverman, J.~M., {et~al.} 2012, The
  Astrophysical Journal Letters, 760, L36

\bibitem[{Romani {et~al.}(2016)Romani, Graham, Filippenko, \&
  Zheng}]{romani2016psr}
Romani, R.~W., Graham, M.~L., Filippenko, A.~V., \& Zheng, W. 2016, The
  Astrophysical Journal, 833, 138

\bibitem[{Romani {et~al.}(2021)Romani, Kandel, Filippenko, Brink, \&
  Zheng}]{romani2021psr}
Romani, R.~W., Kandel, D., Filippenko, A.~V., Brink, T.~G., \& Zheng, W. 2021,
  The Astrophysical Journal Letters, 908, L46

\bibitem[{Romani {et~al.}(2022)Romani, Kandel, Filippenko, Brink, \&
  Zheng}]{romani2022psr}
---. 2022, The Astrophysical Journal Letters, 934, L18

\bibitem[{{Romani} {et~al.}(2022{\natexlab{a}}){Romani}, {Kandel},
  {Filippenko}, {Brink}, \& {Zheng}}]{2022ApJ...934L..17R}
{Romani}, R.~W., {Kandel}, D., {Filippenko}, A.~V., {Brink}, T.~G., \& {Zheng},
  W. 2022{\natexlab{a}}, \apjl, 934, L17, \dodoi{10.3847/2041-8213/ac8007}

\bibitem[{{Romani} \& {Sanchez}(2016)}]{2016ApJ...828....7R}
{Romani}, R.~W., \& {Sanchez}, N. 2016, \apj, 828, 7,
  \dodoi{10.3847/0004-637X/828/1/7}

\bibitem[{{Romani} {et~al.}(2022{\natexlab{b}}){Romani}, {Deller}, {Guillemot},
  {Ding}, {de Vries}, {Parker}, {Zavala}, {Chalumeau}, \&
  {Cognard}}]{2022ApJ...930..101R}
{Romani}, R.~W., {Deller}, A., {Guillemot}, L., {et~al.} 2022{\natexlab{b}},
  \apj, 930, 101, \dodoi{10.3847/1538-4357/ac6263}

\bibitem[{{Sanchez} \& {Romani}(2017)}]{2017ApJ...845...42S}
{Sanchez}, N., \& {Romani}, R.~W. 2017, \apj, 845, 42,
  \dodoi{10.3847/1538-4357/aa7a02}

\bibitem[{Schlafly \& Finkbeiner(2011)}]{schlafly2011measuring}
Schlafly, E.~F., \& Finkbeiner, D.~P. 2011, The Astrophysical Journal, 737, 103

\bibitem[{Schroeder \& Halpern(2014)}]{schroeder2014observations}
Schroeder, J., \& Halpern, J. 2014, The Astrophysical Journal, 793, 78

\bibitem[{Stappers {et~al.}(2001)Stappers, van Kerkwijk, Bell, \&
  Kulkarni}]{stappers2001intrinsic}
Stappers, B., van Kerkwijk, M., Bell, J., \& Kulkarni, S. 2001, The
  Astrophysical Journal Letters, 548, L183

\bibitem[{Stovall {et~al.}(2014)Stovall, Lynch, Ransom, Archibald, Banaszak,
  Biwer, Boyles, Dartez, Day, Ford, {et~al.}}]{stovall2014green}
Stovall, K., Lynch, R., Ransom, S., {et~al.} 2014, The Astrophysical Journal,
  791, 67

\bibitem[{Strader {et~al.}(2019)Strader, Swihart, Chomiuk, Bahramian, Britt,
  Cheung, Dage, Halpern, Li, Mignani, {et~al.}}]{strader2019optical}
Strader, J., Swihart, S., Chomiuk, L., {et~al.} 2019, The Astrophysical
  Journal, 872, 42

\bibitem[{van Kerkwijk {et~al.}(2011)van Kerkwijk, Breton, \&
  Kulkarni}]{van2011evidence}
van Kerkwijk, M., Breton, R., \& Kulkarni, S. 2011, The Astrophysical Journal,
  728, 95

\bibitem[{van Staden \& Antoniadis(2016)}]{van2016active}
van Staden, A.~D., \& Antoniadis, J. 2016, The Astrophysical Journal Letters,
  833, L12

\bibitem[{Voisin {et~al.}(2020)Voisin, Kennedy, Breton, Clark, \&
  Mata-S{\'a}nchez}]{voisin2020model}
Voisin, G., Kennedy, M., Breton, R., Clark, C., \& Mata-S{\'a}nchez, D. 2020,
  \mnras, 499, 1758

\end{thebibliography}
\end{document}